\documentclass[12pt,fleqn]{iopart}
\usepackage{iopams}
\usepackage[abbr,round]{harvard}
\usepackage{graphicx}
\usepackage{bm}
\expandafter\let\csname equation*\endcsname\relax
\expandafter\let\csname endequation*\endcsname\relax
\usepackage{amsmath}

\def\ga{\mathrel{\mathchoice {\vcenter{\offinterlineskip\halign{\hfil
$\displaystyle##$\hfil\cr>\cr\sim\cr}}}
{\vcenter{\offinterlineskip\halign{\hfil$\textstyle##$\hfil\cr
>\cr\sim\cr}}}
{\vcenter{\offinterlineskip\halign{\hfil$\scriptstyle##$\hfil\cr
>\cr\sim\cr}}}
{\vcenter{\offinterlineskip\halign{\hfil$\scriptscriptstyle##$\hfil\cr
>\cr\sim\cr}}}}}

\def\la{\mathrel{\mathchoice {\vcenter{\offinterlineskip\halign{\hfil
$\displaystyle##$\hfil\cr<\cr\sim\cr}}}
{\vcenter{\offinterlineskip\halign{\hfil$\textstyle##$\hfil\cr  
<\cr\sim\cr}}}
{\vcenter{\offinterlineskip\halign{\hfil$\scriptstyle##$\hfil\cr
<\cr\sim\cr}}}
{\vcenter{\offinterlineskip\halign{\hfil$\scriptscriptstyle##$\hfil\cr
<\cr\sim\cr}}}}}

\newcommand{\pof}{Phys. Fluids}  
\newcommand{\gafd}{Geophys.\ Astrophys.\ Fluid\ Dyn.}  
\newcommand{\pre}{Phys. Rev. E}  
\newcommand{\prl}{Phys. Rev. Lett.}  
\newcommand{\jfm}{J. Fluid Mech.}  
\newcommand{\astrona}{Astron. Nachr.}
\newcommand{\prslsa}{Proc. R. Soc. London, Ser. A}

\newcommand{\mur}{\mu_{\rm{r}}}
\newcommand{\rml}{{\rm{Rm}}_{\rm{loc}}}
\renewcommand{\vec}[1]{\mbox{\boldmath$#1$}}

\begin{document}

\title{Magnetic material in mean-field dynamos driven by small scale helical flows}

\author{A. Giesecke, F. Stefani, G. Gerbeth}
\address{Helmholtz-Zentrum Dresden-Rossendorf, 
Institute of Fluid Dynamics,
Department Magnetohydrodynamics, 
01328 Dresden, 
Germany}
\ead{a.giesecke@hzdr.de}

\date{\today}

\begin{abstract}
We perform kinematic simulations of dynamo action driven by a helical
small scale flow of a conducting fluid in order to deduce mean-field
properties of the combined induction action of small scale eddies.  We
examine two different flow patterns in the style of the G.O. Roberts
flow but with a mean vertical component and with internal fixtures
that are modelled by regions with vanishing flow.  These fixtures
represent either rods that lie in the center of individual eddies, or
internal dividing walls that provide a separation of the eddies from
each other.  The fixtures can be made of magnetic material with a
relative permeability larger than one which can alter the dynamo
behavior. The investigations are motivated by the widely unknown
induction effects of the forced helical flow that is used in the core
of liquid sodium cooled fast reactors, and from the key role of soft
iron impellers in the Von-K{\'a}rm{\'a}n-Sodium (VKS) dynamo.

For both examined flow configurations the consideration of magnetic material
within the fluid flow causes a reduction of the critical magnetic Reynolds
number of up to 25\%. The development of the growth-rate in the limit of the
largest achievable permeabilities suggests no further significant
reduction for even larger values of the permeability.

In order to study the dynamo behavior of systems that consist of tens
of thousands of helical cells we resort to the mean-field dynamo
theory \cite{1980mfmd.book.....K} in which the action of the small
scale flow is parameterized in terms of an $\alpha$- and
$\beta$-effect.  We compute the relevant elements of the $\alpha$- and
the $\beta$-tensor using the so called testfield method.  We find a
reasonable agreement between the fully resolved models and the
corresponding mean-field models for wall or rod materials in the
considered range $1\leq \mur \leq 20$.  Our results may be used for
the development of global large scale models with recirculation flow
and realistic boundary conditions.
\end{abstract}


\submitto{\NJP}

\maketitle

\section{Introduction.}

Magnetic fields produced by the flow of a conductive
liquid or plasma can be found in almost all
cosmic objects. In most cases, this does not apply to liquid metal flows
in the laboratory or in industrial applications.
The characteristic properties of these flows -- namely velocity
amplitude, geometric dimension and electrical conductivity -- are
usually not
in the range that allows the occurrence of magnetic self-excitation,
so that an experimental confirmation of the fluid flow driven dynamo effect requires
an enormous effort. 
The aforementioned quantities can be combined into a single,
dimensionless parameter, the magnetic Reynolds number, which is
defined as
${\rm{Rm}}=\mathcal{L}\mathcal{V}/\eta$. Here ${\mathcal{L}}$ is a
typical length scale, $\mathcal{V}$ is a typical velocity amplitude,
and $\eta$ is the  magnetic diffusivity which is the inverse of the
product of vacuum permeability and electrical conductivity
$\eta=(\mu_0\sigma)^{-1}$.  
In cosmic objects, ${\rm{Rm}}$ is typically huge so that one essential precondition 
for the occurrence of dynamo action is fulfilled. 
However, the flow amplitude in terms of ${\rm{Rm}}$ is not the only criterion that
describes the ability of a flow field to provide for dynamo
action, and magnetic self-excitation is also possible at much smaller $\rm{Rm}$ if the fluid
flow has a suitable structure.

\begin{figure}[h!]
\hspace*{4cm}
\includegraphics[width=7.0cm]{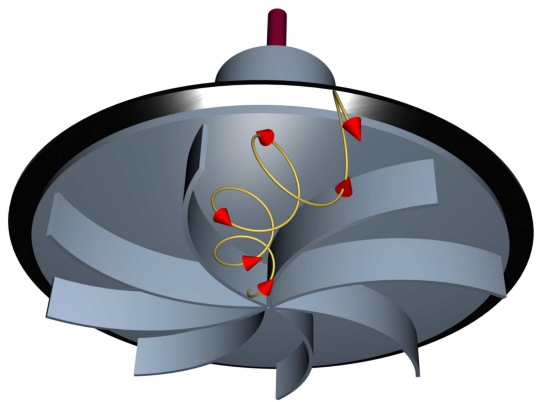}
\caption{Sketch of an individual impeller that drives the
  von-K{\'a}rm{\'a}n-like flow in the VKS dynamo. The yellow
  streamline denotes the assumed helical flow between adjacent
  blades. For dynamo action to occur, both the disk and the blades
  must be made of soft-iron.}
\label{fig::vks}
\end{figure}

Appropriate flows have been utilized, for example, in the three
successful fluid flow driven dynamo experiments, the Riga dynamo
\cite{2000PhRvL..84.4365G}, the Karlsruhe dynamo
\cite{2001PhFl...13..561S}, and the Von-K{\'a}rm{\'a}n-Sodium (VKS)
dynamo \cite{2007PhRvL..98d4502M}.  Both, Riga dynamo and Karlsruhe
dynamo, were based on a screw-like flow pattern, utilizing the fact
that helicity is conducive for the occurrence of dynamo action
\cite{ISI:000080074100003}.  The role of helicity is less obvious for
the VKS dynamo with a flow of liquid sodium being driven by two
counter-rotating impellers.  It has long been known that the mean flow
generated by this forcing is suited to drive a dynamo at comparatively
low ${\rm{Rm}}$ \cite{1989RSPSA.425..407D}. 
However, in the experimental implementation at the VKS dynamo, the motor
power available to drive the flow is not sufficient to overcome the threshold
for the equatorial dipole mode with an azimuthal wavenumber
$m=1$. Surprisingly, dynamo action of the axisymmetric dipole mode has yet 
been found at a rather low magnetic Reynolds number ${\rm{Rm}}\approx 32$ but
only if the entire flow driving system, consisting of a disk and eight bended
blades (figure~\ref{fig::vks}), is made of soft-iron with a relative
permeability in the order of $\mur\approx 60$
\cite{2010NJPh...12c3006V,2013PhRvE..88a3002M}. 
A possible explanation for this observation requires the combined
effects of the magnetic properties of the soft iron disks
\cite{2012NJPh...14e3005G}, and helical radial outflows assumed in the
vicinity of the impellers between adjacent blades
\cite{2007GApFD.101..289P}. These non-axisymmetric distortions of the
mean flow can be parameterized by an $\alpha$-effect
(figure~\ref{fig::vks}), but so far existing mean-field models of the
VKS dynamo are only of limited significance due to a lack of knowledge
about the $\alpha$-effect and its interaction with the magnetic
material of the impeller systems
\cite{2010GApFD.104..249G,2010PhRvL.104d4503G}.

\begin{figure}[h!]
\vspace*{0.3cm}
\includegraphics[width=16cm]{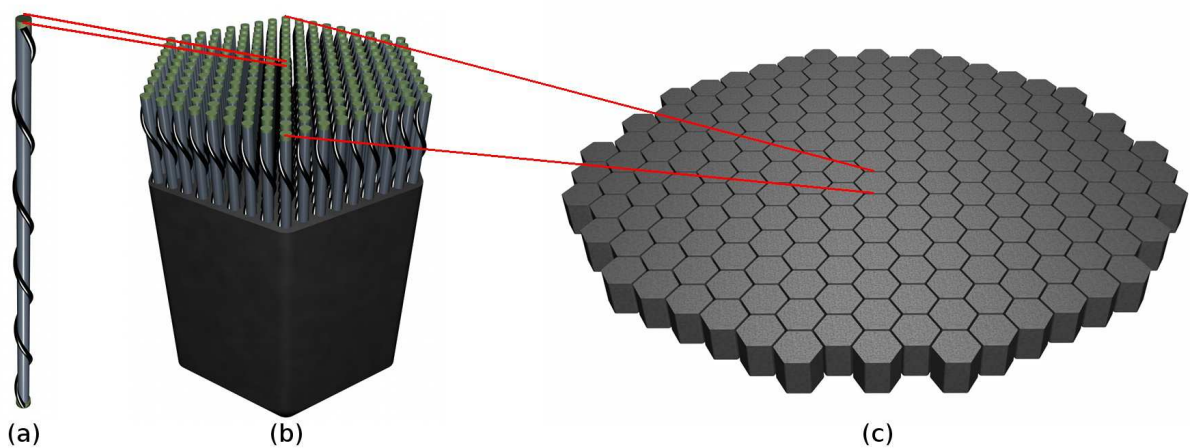}
\caption{Idealized composition of the core of a sodium fast reactor;
  from left to right: 
  (a) nuclear fuel rod surrounded by a helical
  shaped spacer forcing the flow on a helical path; 
  (b) assembly of bundled fuel rods; 
  (c) array of assemblies, forming the core of a liquid metal cooled
  fast reactor. Note that the figure shows an idealized
  system. In real systems, there are still additional elements with
  breeding material and control rods.}
\label{fig::lmcfbr}
\end{figure}

Besides of the relevance for understanding the fundamental physics of
geo- and astrophysical magnetic fields, a complementary argument for
the development and construction of dynamo experiments originated from
considerations on the safe operation of sodium cooled fast reactors
\cite{mhdhistory2}.  Dynamo action in the cooling system of a sodium
fast reactor would likely be dangerous, because the self-induced
magnetic field backreacts on the flow according to Lenz's law.  This
backreaction might cause an inhomogeneous flow breaking or a pressure
drop in the pipe system so that the efficient cooling of the reactor
core would be hampered with unknown consequences for the safety of the
reactor.  The occurrence of dynamo action in a sodium fast reactor can
not be excluded a priori because the flow in the core has a
sufficiently large flow rate, and the appropriate geometry.  In the
very core of the reactors the fluid flow is governed by screw-like
shaped wires that are wrapped around individual nuclear fuel rods thus
forcing the flow to follow a helical path around each rod
(figure~\ref{fig::lmcfbr}a).
These fuel rods are bundled into so called assemblies which may
consist of up to a few hundreds of fuel rods
(figure~\ref{fig::lmcfbr}b), and the whole reactor core is composed of
a few hundreds of these assemblies
(figure~\ref{fig::lmcfbr}c)\footnote[7]{For instance, the core of the
French reactor {\it{Superphenix}} contained 364 fuel assemblies,
each comprising 271 fuel rods.}.  In operation, this setup is
flushed with liquid sodium thereby forming a helical flow field that
is reminiscent of the flow used in the Karlsruhe dynamo (except the
mean vertical component).

Actually, early estimations by \citename{bevir} \citeyear{bevir} and
\citename{pierson} \citeyear{pierson} as well as more recent
experimental and numerical studies
\cite{1995maghyd...31.4..382P,1999JFM...382..137P,2000JFM...403..263A}
show no conclusive evidence for the occurrence of dynamo action in the
core of a fast reactor.  On the other hand it has been argued by
\citename{thesis_soto} \citeyear{thesis_soto} that the parameter
regime reached by the French fast breeder reactor {\it{Superphenix}}
is well within the range that allows for dynamo action if some
magnetic material is introduced into the container
\citeaffixed{thesis_soto}{see p. 104, Fig III.32 in}.  So far, the
problem of magnetic material in the core of a sodium fast reactor is
merely academic, because state of the art reactors mainly utilize
austenitic steels inside the core.  However, in recent years, the
application of {\it{Oxide Dispersion Strengthened}} (ODS)
ferritic/martensitic alloys with a relative permeability $\mur \gg 1$
has increasingly been discussed because these alloys have a lower
sensitivity for nuclear radiation \cite{2012JNuM..428....6D}.

The dramatic influence of magnetic properties on the induction
process, as observed at the VKS dynamo, motivated the present study,
in which we examine complex interactions of helical flow fields with
magnetic internals.  Since the flow conditions in a sodium fast
reactor are far too complex to be modeled in direct numerical
simulations, we resort to the mean-field dynamo theory, which allows
the development of models that are numerically much easier to handle.
In order to consider the specific effects of a spatially varying
permeability distribution we extend the original mean-field concept to
the case of non-uniform material properties.  The extension is
straightforward and allows to take into account complex periodic
patterns with magnetic properties in terms of standard mean-field
coefficients like the $\alpha$- and $\beta$-effect.  For the
estimation of the mean-field coefficients we perform kinematic
simulations of electromagnetic induction generated by idealized
helical flow fields that are reminiscent of the conditions in sodium
fast reactors.  We consider two paradigmatic configurations with
either a helical flow subdivided by internal walls, or a flow
following a helical path around solid rods, respectively.  The first
model follows the heuristic approach of \citename{pierson}
\citeyear{pierson} in which the screw-like vortex represents the mean
flow within an assembly of nuclear fuel rods.  In a very broad sense,
this model can also serve as an approach for the flow field between
the blades in the VKS dynamo.  The second model goes back to the work
of \citename{2002NPGeo...9..171R}
\citeyear{2002NPGeo...9..171R,2002MHD....38...41R} on the kinematic
theory of the Karlsruhe dynamo.  In the present study, in which we
assume a vertical mean flow, this type of flow field is suited to the
conditions within an assembly of fuel rods in the core of a sodium
fast reactor.

We start with the analysis of the induction action of the fully
resolved velocity field, from which we determine the mean-field
coefficients using the testfield method
\cite{2005AN....326..245S,2007GApFD.101...81S}.  In a second step we
use the $\alpha$- and $\beta$-coefficients as an input for mean field
dynamo simulations in order to prove that mean-field models are
capable to reproduce the growth-rate and principle field structure of
the fully resolved model by requiring much less computational efforts.
For flow systems comprising a total of some tens of thousands of
individual helical cells (figure~\ref{fig::lmcfbr}), the use of a
well-proven mean-field method is considered the only viable way to
study dynamo problems. The present paper is mainly intended to
establish and validate the necessary methodology. The possible
application to specific reactor cores will need much more information
on geometric details and material properties, and must therefore be
left for future work.

\section{Mean-field dynamo theory and the testfield method}

\subsection{Outline of mean-field theory}

In the following, the magnetic flux density is denoted by $\vec{B}$
and the velocity field by $\vec{U}$. The magnetic diffusivity is
defined by $\eta=(\mu\sigma)^{-1}$ with the electrical conductivity
$\sigma$ and the permeability $\mu$ which are assumed, for the moment,
to be constant.  The temporal development of the magnetic flux density
in the presence of an electrically conductive liquid that moves
according to the velocity field $\vec{U}$ is determined by the
induction equation:
\begin{equation}
\frac{\partial\vec{B}}{\partial t}=\nabla\times\left(\vec{U}\times\vec{B}
-\eta\nabla\times{\vec{B}}\right).\label{eq::ind}
\end{equation}
Additionally, $\vec{B}$ must obey the divergence-free condition,
$\nabla\cdot\vec{B}=0$.  In case of a prescribed (stationary) velocity
field, equation ~(\ref{eq::ind}) is a linear problem which, in
principle, can be solved with the Ansatz
\begin{equation}
\vec{B}(\vec{r}, t) =\vec{B}_0(\vec{r})e^{\lambda t}.  
\end{equation}
In general $\lambda$ is a complex quantity $\lambda=\kappa+i\omega$
where $\kappa$ denotes the growth-rate and $\omega$ denotes an
oscillation- or drift-frequency. A dynamo solution is obtained if the
magnetic field amplitude $|\vec{B}|$ grows exponentially $\propto
e^{\kappa t}$ with a growth-rate $\kappa>0$.

Even though the linear approach is a severe simplification that
neglects the backreaction of the field on the flow, equation
(\ref{eq::ind}) can be solved analytically only for very few cases. In
particular for complicated velocity fields with small scale
structures, equation (\ref{eq::ind}) must be solved numerically.  A
possibility to draw further conclusions on the ability of a velocity
field to drive a dynamo is provided by the mean-field dynamo theory
developed by \citename{1980mfmd.book.....K}
\citeyear{1980mfmd.book.....K}.  The mean-field dynamo theory
essentially deals with the behavior of the large scale field and
treats the induction effects of a small scale flow in terms of the so
called $\alpha$-effect. The basic principle of the mean-field approach
is a splitting of magnetic field and velocity field assuming that the
properties of the whole system can be described essentially by two
scales, a mean, large scale part ($\overline{\vec{B}}$ and
$\overline{\vec{U}}$) and a small scale fluctuation ($\vec{b}$ and
$\vec{u}$):
\begin{eqnarray}
\vec{B}&=&\overline{\vec{B}}+\vec{b},\label{eq::split_magfield}\\
\vec{U}&=&\overline{\vec{U}}+\vec{u}.\label{eq::split_velfield}
\end{eqnarray}
Inserting (\ref{eq::split_magfield}) and (\ref{eq::split_velfield})
into (\ref{eq::ind}) yields an induction equation for the mean-field
$\overline{\vec{B}}$:
\begin{equation}
\frac{\partial\overline{\vec{B}}}{\partial t}=\nabla\times\left(\overline{\vec{U}}\times\overline{\vec{B}}
+\overline{\vec{u}\times\vec{b}}-\eta\nabla\times\overline{\vec{B}}\right), 
\label{eq::b_mean}
\end{equation}
while the induction equation for the corresponding small scale field
$\vec{b}$ reads: 
\begin{equation}
\frac{\partial \vec{b}}{\partial t}=
\nabla\times\left(\overline{\vec{U}}\!\times\!\vec{b}
+\vec{u}\!\times\!\overline{\vec{B}}
+\left(\vec{u}\!\times\!\vec{b}-\overline{\vec{u}\!\times\!\vec{b}}\right)
-\eta\nabla\!\times\!\vec{b}\right).
\label{eq::b_small}
\end{equation}
Furthermore, the mean-field as well as the small scale field must obey
$\nabla\cdot\overline{\vec{B}}\!=\!0$ and $\nabla\cdot\vec{b}\!=\!0$.
The mean-field induction equation~(\ref{eq::b_mean}) contains an
additional source term,
$\boldmath{\mathcal{E}{\vec{}}}\!=\!\overline{\vec{u}\!\times\!\vec{b}}$,
called the mean electromotive force (EMF).  In the kinematic
approximation, $\boldmath{\mathcal{E}{\vec{}}}$ is linear and
homogeneous in $\overline{\vec{B}}$, and, under the assumption that
the variations of $\overline{\vec{B}}$ around a given point are small,
$\boldmath{\mathcal{E}\vec{}}$ can be represented by the first terms
of a Taylor expansion:
\begin{equation}
{\mathcal{E}}_i=\alpha_{ij} B_j
+\beta_{ijk}\frac{\partial B_j}{\partial x_k}.\label{eq::emf}
\end{equation}
Here $\alpha_{ij}$ and $\beta_{ijk}$ are tensors of second and third
rank, respectively. The diagonal components $\alpha_{ii}$ give rise
to an electromotive force parallel to the mean magnetic field and therefore may
be responsible for dynamo action. For isotropic turbulence, the
contribution proportional to the mean-field gradients simplifies to
$\beta\epsilon_{ijk}\partial B_j/\partial x_k$ (with the Levi-Civita
tensor $\epsilon_{ijk}$), so that this term behaves similar to a
diffusive contribution.  However, in our setup we have a strong
anisotropy between vertical and horizontal coordinates, so that we
refer to another expression for the electromotive force, that is based
on elementary symmetry properties of flow and field
\cite{1980mfmd.book.....K}:
\begin{eqnarray}
\vec{\mathcal{E}}&=&-\alpha_{\perp}\overline{\vec{B}}
-(\alpha_{\parallel}\!-\!\alpha_{\perp})(\hat{\vec{z}}\!\cdot\!\overline{\vec{B}})\hat{\vec{z}}
-\gamma\hat{\vec{z}}\!\times\!\overline{\vec{B}}
-\beta_{\perp}\nabla\!\times\!\overline{\vec{B}}
-(\beta_{\parallel}\!-\!\beta_{\perp})(\hat{\vec{z}}\!\cdot\!(\nabla\!\times\!\overline{\vec{B}}))\hat{\vec{z}}\nonumber\\
&&-\widetilde{\beta}\hat{\vec{z}}\!\times\!(\nabla(\hat{\vec{z}}\!\cdot\!\overline{\vec{B}})+(\hat{\vec{z}}\!\cdot\!\nabla)\overline{\vec{B}})
-\delta_1\nabla(\hat{\vec{z}}\!\cdot\!\overline{\vec{B}})-\delta_2(\hat{\vec{z}}\!\cdot\!\nabla)\overline{\vec{B}}
-\delta_3(\hat{\vec{z}}\!\cdot\!\nabla(\hat{\vec{z}}\!\cdot\!\overline{\vec{B}}))\hat{\vec{z}}.\label{eq::emf_general}
\end{eqnarray}
Here, a mean flow is assumed along the vertical direction which is
labeled by $\hat{\vec{z}}$ in a Cartesian system.  The subscript
$\parallel$ denotes quantities that are parallel to this vertical
direction whereas the subscript $\perp$ denotes quantities that are
oriented in the horizontal plane ($xy$-plane).  In
equation~(\ref{eq::emf_general}), $\alpha_{\perp}$ and
$\alpha_{\parallel}$ give rise to a current parallel to the mean
magnetic field and, hence, can be responsible for dynamo action.
These coefficients correspond to the diagonal elements of the
$\alpha$-tensor, and anisotropic effects arising from properties of
the small scale velocity field result in different contributions from
the horizontal part $\alpha_{\perp}$ (that generates a current in the
$xy$-plane) and the vertical part $\alpha_{\parallel}$ (that generates
a current along the $z$-axis).  In the same way, $\beta_{\perp}$ and
$\beta_{\parallel}$ can be interpreted as anisotropic contributions
to the magnetic diffusivity.  The coefficient $\gamma$ is related to
the antisymmetric part of the $\alpha$-tensor and describes an
additional advection of the mean-field in the direction of the mean
flow.  The remaining coefficients $\widetilde{\beta}$ and $\delta_i$
are related to the gradient tensor of the magnetic field and have no
simple analogy.  A more detailed derivation of
equation~(\ref{eq::emf_general}) and a discussion about the mean-field
coefficients $\alpha_{\perp}, \alpha_{\parallel},
\beta_{\perp},\beta_{\parallel},\widetilde{\beta},\delta_{i}$ are
given in the textbook of~\citename{1980mfmd.book.....K}
\citeyear{1980mfmd.book.....K}.

In the following, we only consider flow fields that do not depend on
$z$ and that are periodic in the $xy$-plane. All mean quantities are
defined as horizontal averages, i.e., they do not depend on $x$ or
$y$. Consequently, most of the coefficients and all terms proportional
to mean-field gradients in $x$ and $y$ vanish so that
(\ref{eq::emf_general}) can be significantly simplified:
\begin{equation}
\vec{\mathcal{E}}=-\alpha_{\perp}(\overline{\vec{B}}-(\hat{\vec{z}}\cdot\overline{\vec{B}})\hat{\vec{z}})
-(\beta_{\perp}+\widetilde{\beta})\hat{\vec{z}}\times\frac{d\overline{\vec{B}}}{dz}
-\gamma{\hat{\vec{z}}}\times\overline{\vec{B}}-\delta_2\frac{d\overline{\vec{B}}}{dz}.
\label{eq::simp_emf}
\end{equation}
Note, that due to the constant velocity along $z$ and the vanishing
horizontal derivatives of the mean-field, all contributions labeled
with $\parallel$ can be dropped and only two terms $\propto
\partial\overline{\vec{B}}/\partial z$ survive.  Furthermore, the
effects corresponding to $\beta_{\perp}$ and $\widetilde{\beta}$
cannot be distinguished any more and are subsumed into one common
coefficient $\beta=\beta_{\perp}+\widetilde{\beta}$
\cite{2003PhRvE..67b6401R}.

The tensor coefficients appearing in (\ref{eq::emf}) can be
related to the more descriptive notation used in
(\ref{eq::simp_emf}) giving the following relations:
\begin{eqnarray}
\alpha_{xx} = &\quad\alpha_{yy}  & =  -\alpha_{\perp},\nonumber\\
\alpha_{xy} = &-\alpha_{yx} & =  \quad\gamma,\label{eq::isotropy}\\
\beta_{xyz} = & -\beta_{yxz} & =  -(\beta_{\perp}+\widetilde{\beta})=-\beta,\nonumber\\
\beta_{xxz} = & \quad\beta_{yyz}  & =  -\delta_2.\nonumber
\end{eqnarray}
These relations reflect the horizontal isotropy in our models and
allow a simplification of the problem since only four
coefficients must be determined in order to establish a
consistent mean-field model.

\subsection{Testfield method}

The test field method developed in \citeasnoun{2005AN....326..245S}
provides a powerful tool to compute the coefficients $\alpha_{ij}$ and
$\beta_{ijk}$ from different realizations of the electromotive force
that are obtained from externally applied, linearly independent
mean-fields.  Here, we restrict ourselves to the kinematic case with a
stationary velocity field although the method can also be applied to
fully non-linear magnetohydrodynamic systems where $\vec{U}$ is computed by solving
the Navier-Stokes equation.

The fluctuating velocity field is
computed from the full velocity field $\vec{U}$ by 
\begin{equation}
\vec{u}=\vec{U}-\overline{\vec{U}}
\end{equation}
with $\overline{\vec{U}}$ being the horizontal average of $\vec{U}$.
The small scale magnetic field $\vec{b}$ is computed numerically by
solving equation (\ref{eq::b_small}) with $\overline{\vec{B}}$ defined
as an external steady field, the so called testfield.  Then the
electromotive force is computed directly by correlating small scale
flow with the small scale field and subsequently performing a
horizontal averaging:
$\boldmath{\mathcal{E}}=\overline{\vec{u}\times\vec{b}}$.  The
combination of different realizations of
$\boldmath{\mathcal{E}}\vec{}$ obtained from different, linearly
independent testfields with (\ref{eq::emf}) yields a linear system of
equations whose solution gives the desired mean-field coefficients.
In principle, only minor preconditions for the testfields must be
considered.  In order to calculate mean-field coefficients that are
consistent with the structure of the large scale field obtained
from the fully resolved model it is necessary to consider the scale
dependence of the mean-field coefficients \cite{2008A&A...482..739B}.
Around the onset of dynamo action the vertical dependence of the large
scale field in our systems is $\propto \cos(z)$ (or $\propto
\sin(z)$) which is exactly the vertical structure that we imply to
the testfields. Because of the horizontal isotropy of our system, we
define two testfields oriented in the horizontal plane parallel to
$\hat{\vec{y}}$:
\begin{equation}
\overline{\vec{B}}_1  =  \cos(z)\hat{\vec{y}} \mbox{ and }
\overline{\vec{B}}_2  =  \sin(z)\hat{\vec{y}}.
\label{eq::testfields}
\end{equation}
With this definition we obtain four equations with four unknown mean
field coefficients $\alpha_{yy},\alpha_{xy},\beta_{yyz},
\beta_{xyz}$ which read: 
\begin{eqnarray}
{\alpha}_{yy}&=&\overline{\mathcal{E}}_{1,y}(z)
\cos(z)+\overline{\mathcal{E}}_{2,y}(z)\sin(z),\\
{\beta}_{xyz}&=&-\left(\overline{\mathcal{E}}_{1,x}(z)
\sin(z)-\overline{\mathcal{E}}_{2,x}(z)\cos(z)\right),\\
\alpha_{xy} & = &\overline{\mathcal{E}}_{1,x}(z)
\cos(z)+\overline{\mathcal{E}}_{2,x}(z)\sin(z),\\
\beta_{yyz} & = &-\left(\overline{\mathcal{E}}_{1,y}(z)
\sin(z)-\overline{\mathcal{E}}_{2,y}(z)\cos(z)\right).
\end{eqnarray}
Here $\mathcal{E}_{1,2,x,y}$ denote the horizontal components of the
electromotive force obtained with the testfields
$\overline{\vec{B}}_{1,2}$.  The remaining coefficients $\alpha_{xx},
\alpha_{yx}, \beta_{yxz}$ and $\beta_{xxz}$ can, in principle, be
calculated with similar equations involving
$\boldmath{\mathcal{E}}{\vec{}}$ obtained from
$\overline{\vec{B}}_{3}=\cos(z)\hat{\vec{x}}$ and
$\overline{\vec{B}}_{4}=\sin(z)\hat{\vec{x}}$ which requires the
numerical solution of two further partial differential equations for
the corresponding $\vec{b}$.  For test purposes we have additionally
performed these calculations and verified that the isotropy conditions
given by (\ref{eq::isotropy}) are met in the simulations.

\section{Flow models and permeability distribution}

\subsection{Velocity field}

\begin{figure}[h!]
\includegraphics[width=10cm]{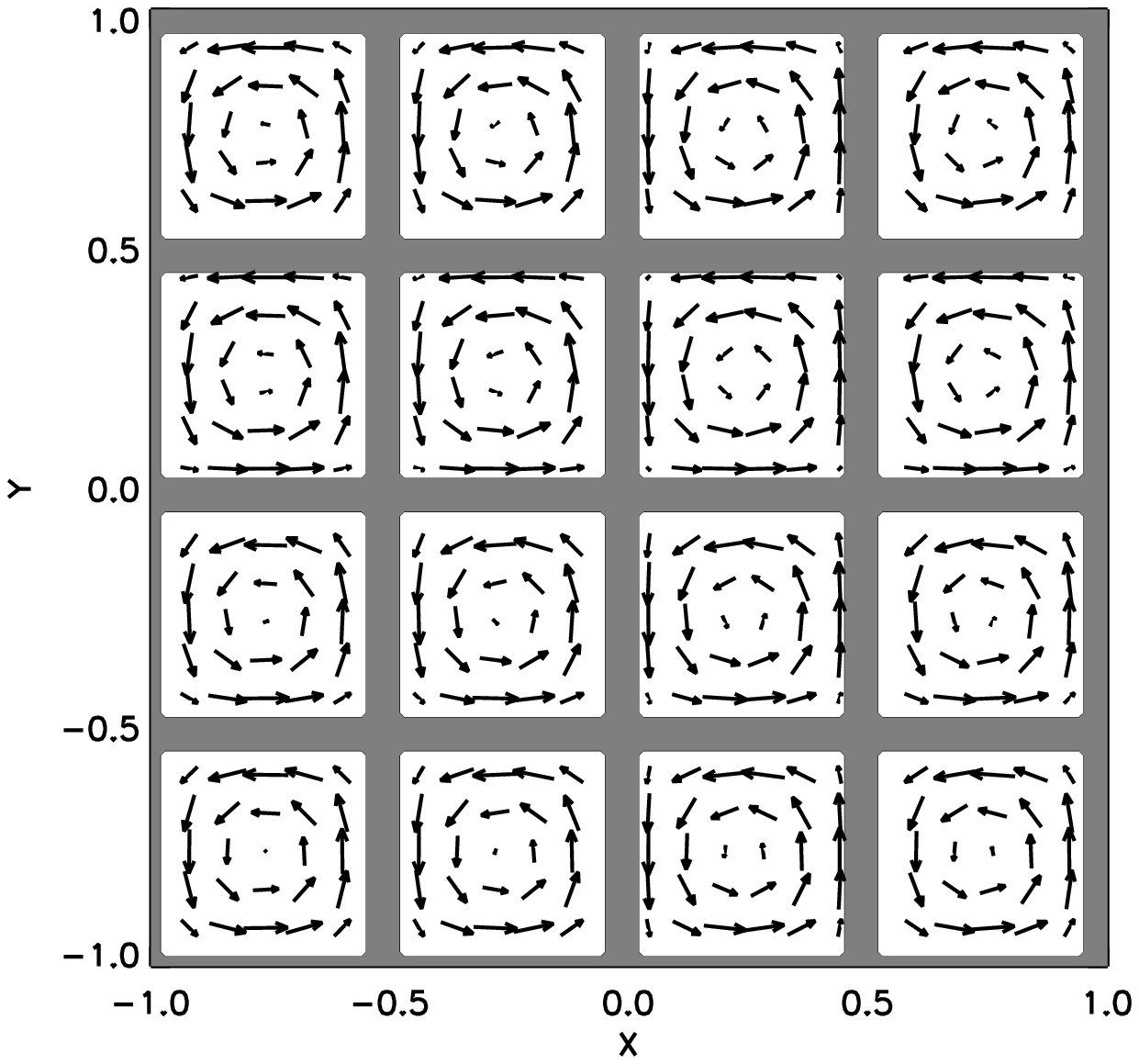}
\nolinebreak[4!]
\hspace*{-2cm}
\nolinebreak[4!]
\includegraphics[width=10cm]{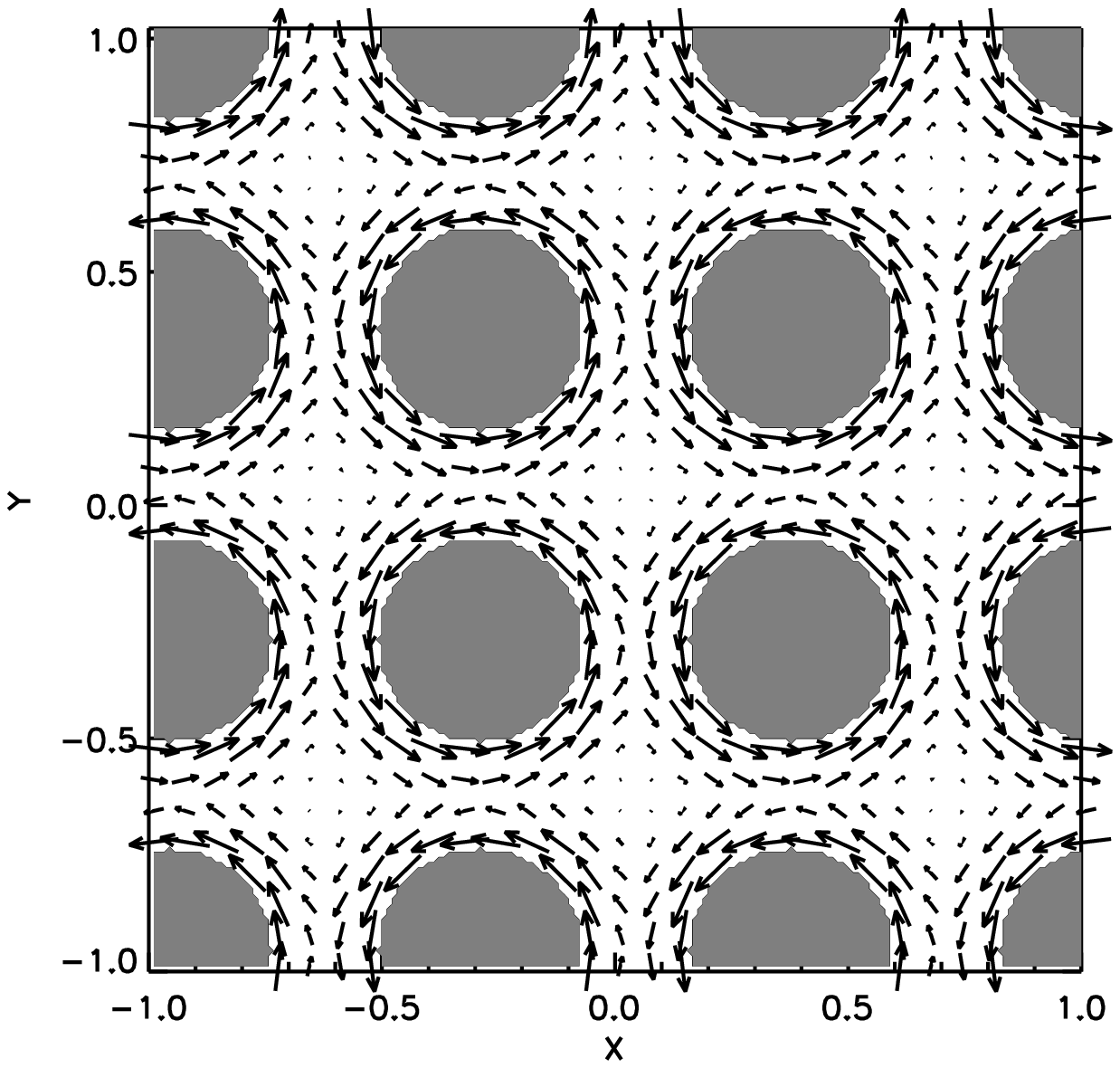}
\caption{Flow pattern for model A (left panel) and model B (right
  panel). The gray shaded regions represent walls (model A, left) or
  rods (model B, right) and the arrows denote the horizontal flow. The
  vertical flow is constant 
  within each cell and vanishes in the wall/rod regions. The
  fluid permeability is $\mur=1$ and the wall/rod permeability is varied
  in the range $1\le \mur\le 20$.}\label{fig::flowfield}
\end{figure}

In the present study we examine two different flow models: In model A
we assume a flow consisting of various helical eddies that are
separated by walls (left panel in figure~\ref{fig::flowfield}).  This
flow definition resembles the Roberts-flow
\cite{1970RSPTA.266..535R,1972RSPTA.271..411R} but comprises a
separating region between each cell quite similar to the model
examined by \citename{2005AN....326..250S}
\citeyear{2005AN....326..250S}.  In contrast to the Roberts-flow, the
flow in our model has the same orientation (left-handed) in every
cell. However, in combination with a uniform vertical flow, each cell
provides the same helicity as it is also the case for the Roberts
flow.  We further allow for a variation of the relative permeability
assuming that magnetic material is used to guide the flow along the
vertical direction.  Following the idea of \citeasnoun{pierson}, the
helical flow within one cell represents the mean flow within one
assembly of nuclear fuel rods ignoring the even smaller scale flow
around individual rods.

The flow with amplitude $u_0$ in one individual cell of size $D$ is given by 
\begin{eqnarray}
u_x & = & \quad\! u_0\sin\left(\pi\frac{x-x_0}{D}\right)\cos\left(\pi\frac{y-y_0}{D}\right),\nonumber\\
u_y & = & -u_0\cos\left(\pi\frac{x-x_0}{D}\right)\sin\left(\pi\frac{y-y_0}{D}\right),\label{eq::flowA}\\
u_z & = &  \quad\! u_0,\nonumber 
\end{eqnarray}
where $x_0$ and $y_0$ represent the coordinates of the cell center in
the horizontal plane.  
The total velocity field is a superimposition of $N$
cells (each using~(\ref{eq::flowA})) which additionally considers the wall regions by setting
$v_x=v_y=v_z=0$ there.
The thickness of the walls is defined as
$d=2/(6\sqrt{N})$ with $N$ the number of helical cells.  The
definition of the wall thickness ensures that the relation of cell
size to wall thickness is constant when increasing the number of
cells.  The specific value is chosen so that the number of grid
points representing a wall is sufficient to numerically resolve the
effects of the permeability transition of the fluid-wall interface.
We used four different realizations with $N=4, 16, 64$ and $256$ cells
arranged in a squared pattern, however, most simulations have been
performed using the setup shown in the left panel of
figure~\ref{fig::flowfield} where $16$ cells in a horizontal plane are
displayed.

The second approach (model B, see right panel in
figure~\ref{fig::flowfield}) uses a more detailed picture of the flow
conditions within a single assembly.  The model is based on the so
called spin generator flow that has been utilized for the simulation
of the Karlsruhe Dynamo
\cite{2002NPGeo...9..171R,2002MHD....38...41R,2003PhRvE..67b6401R}.  A
detailed numerical model of the flow in a hexagonal assembly
consisting of seven fuel rods including a wire wrap surrounding each
rod can be found in \citename{gajapathy} \citeyear{gajapathy} where
the Navier-Stokes equation is solved numerically and turbulence
effects are included in terms of a standard $k-\epsilon$ model.  Here,
we use a simplified flow field roughly in accordance with the model of
\citename{2002MHD....38...41R}
\citeyear{2002MHD....38...41R,2002NPGeo...9..171R} by assuming a
circular flow around a central rod superimposed with a constant
vertical flow.  The flow around a single rod is defined as
\begin{eqnarray}
u_x & = &-\frac{1}{2}u_0\frac{y-y_0}{\sqrt{(x-x_0)^2+(y-y_0)^2}}
         \left(1+\cos\!\left(\pi\frac{\sqrt{(x-x_0)^2+(y-y_0)^2}-R}{D}\right)\!\right)\nonumber\\
u_y & = &\frac{1}{2}u_0\frac{x-x_0}{\sqrt{(x-x_0)^2+(y-y_0)^2}}
         \left(1+\cos\!\left(\pi\frac{\sqrt{(x-x_0)^2+(y-y_0)^2}-R}{D}\right)\!\right)\label{eq::flowB}\\
u_z & = & u_0\nonumber
\end{eqnarray}
where $(x_0,y_0)$ is the center of a rod, $R$ is the radius of the
rod  and $D$ is the distance between two adjacent rods
(see right panel of figure~\ref{fig::flowfield}).
We have performed simulations with $9$ and $25$ rods regularly
distributed in the horizontal plane.

Note, that all helical flow cells in both models are left-handed, so
that the helicity provided by each cell has the same sign.
Furthermore, the global dimensions of the computational domain remain
the same, independent of the number of cells, so that an increasing
number of cells or rods goes along with a smaller scale of the
fluctuating flow component and, thus, an increased separation between
large scale and small scale flow.  Horizontal isotropy is preserved by
applying a quadratical configuration (identical linear extensions and
identical resolution) and periodic boundaries.  
The vertical extent of the computational domain is $z\in[0;2\pi]$
with periodic boundaries as well.

The Cartesian geometry
is different from the hexagonal pattern of realistic
assemblies. However, we believe that for the development of the
methodology, the numerically much easier to handle Cartesian geometry
is more advantageous without exhibiting excessive deviations from the
realistic case.

In order to characterize the amplitude of the flow we define a local
magnetic Reynolds number that is based on the flow amplitude $u_0$,
the ``normal'' magnetic diffusivity $\eta=(\mu_0\sigma)^{-1}$ and the
size $D$ of a single eddy (model A) or the distance between two
adjacent rods (model B):
\begin{equation}
{\rm{Rm}}_{\rm{loc}}=\frac{u_0 D}{\eta}.
\end{equation}

\subsection{Permeability distribution}

The standard mean-field approach developed in
\citeasnoun{1980mfmd.book.....K} is not intended to consider a
spatially varying (``fluctuating'') permeability which can easily be
seen taking the case ${\rm{Rm}}=0$. Then the EMF must vanish,
${\boldmath{\mathcal{E}}{\vec{}}}=0$ (since $\vec{u}=0$) and thus all
mean-field coefficients vanish independently from the actual
distribution of $\mur$.  Our modification starts with the induction
equation with a non-uniform permeability distribution
$\mur=\mur(\vec{r})$, which reads
\begin{equation}
\frac{\partial\vec{B}}{\partial t} =
\nabla\times\left(\vec{U}\times\vec{B}-\eta\nabla\times\frac{\vec{B}}{\mur}\right).\label{eq::ind_with_mur} 
\end{equation}
Using standard vector relations and $\nabla\cdot\overline{\vec{B}}=0$
we rewrite (\ref{eq::ind_with_mur}) in the form  
\begin{equation}
\frac{\partial\vec{B}}{\partial t} =
\nabla\times\left(\vec{U}\times\vec{B}+{\widetilde{\eta}}\frac{\nabla\mur}{\mur}\times{\vec{B}}
-\widetilde{{\eta}}\,\nabla\times\vec{B}\right)\label{eq::ind_with_mur_2} 
\end{equation}
with $\widetilde{\eta}=\widetilde{\eta}(\vec{r})=\eta/\mur(\vec{r})$.
The modified induction equation (\ref{eq::ind_with_mur_2}) exhibits an
additional, not necessarily divergence-free, velocity-like term,
sometimes called {\it{paramagnetic pumping}}
\cite{2003PhRvE..67e6309D}:
\begin{equation}
\widetilde{\vec{u}}(\vec{r})=\widetilde{\eta}\frac{\nabla\mur}{\mur(\vec{r})}.\label{eq::parapump}
\end{equation}
We define a modified velocity field
$\widetilde{\vec{U}}=\vec{U}+\widetilde{\vec{u}}$ which is now the
velocity field that has to be split up into mean part and fluctuating
part when applied in the testfield method.  Note, that the
introduction of the pumping velocity $\widetilde{\vec{u}}$ provides a
non-vanishing fluctuating velocity contribution even in case of a
vanishing fluid flow (i.e. when $u_0=0$).  In our model, we first
define a permeability distribution, from which we compute the
corresponding pumping velocity $\widetilde{\vec{u}}$ using a simple
finite difference discretisation.  In the fluid regions (where
$u_z\neq 0$), the permeability distribution takes the value $\mur=1$,
and $\mur$ is set to a fixed value $\neq 1$ in the remaining regions
(where $u_z=0$, indicated by the grey shaded areas in
figure~\ref{fig::flowfield}).  In order to avoid the discontinuity at
the fluid-solid body transition, which would lead to an amplitude for
the pumping velocity that depends on the grid-resolution, we smoothed
the discontinuity at the fluid-solid body interface by assuming some
sinusoidal distribution with a fixed length-scale that is independent
of the grid resolution.

\section{Results}
In this section, we will apply the test field method to
the two geometric models A and B, first without and then
with consideration of magnetic materials. In each case we
will validate the correspondence of the dynamo action of the
fully resolved and the derived mean-field models.

\subsection{Homogeneous case ($\mur=1$)}

\begin{figure}[h!]
\hspace*{1.5cm}
\includegraphics[width=6cm]{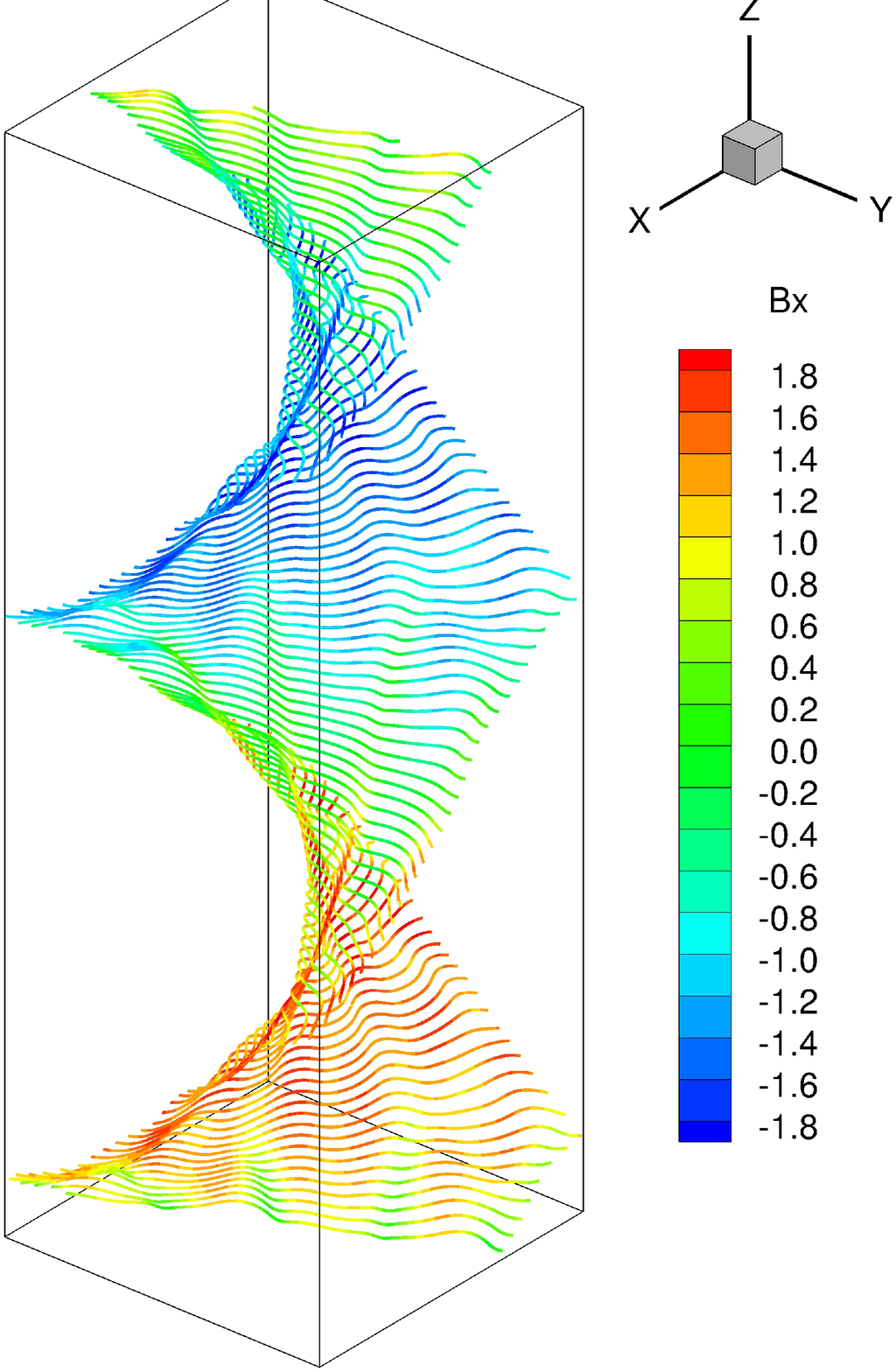}
\hspace*{1.0cm}
\includegraphics[width=6cm]{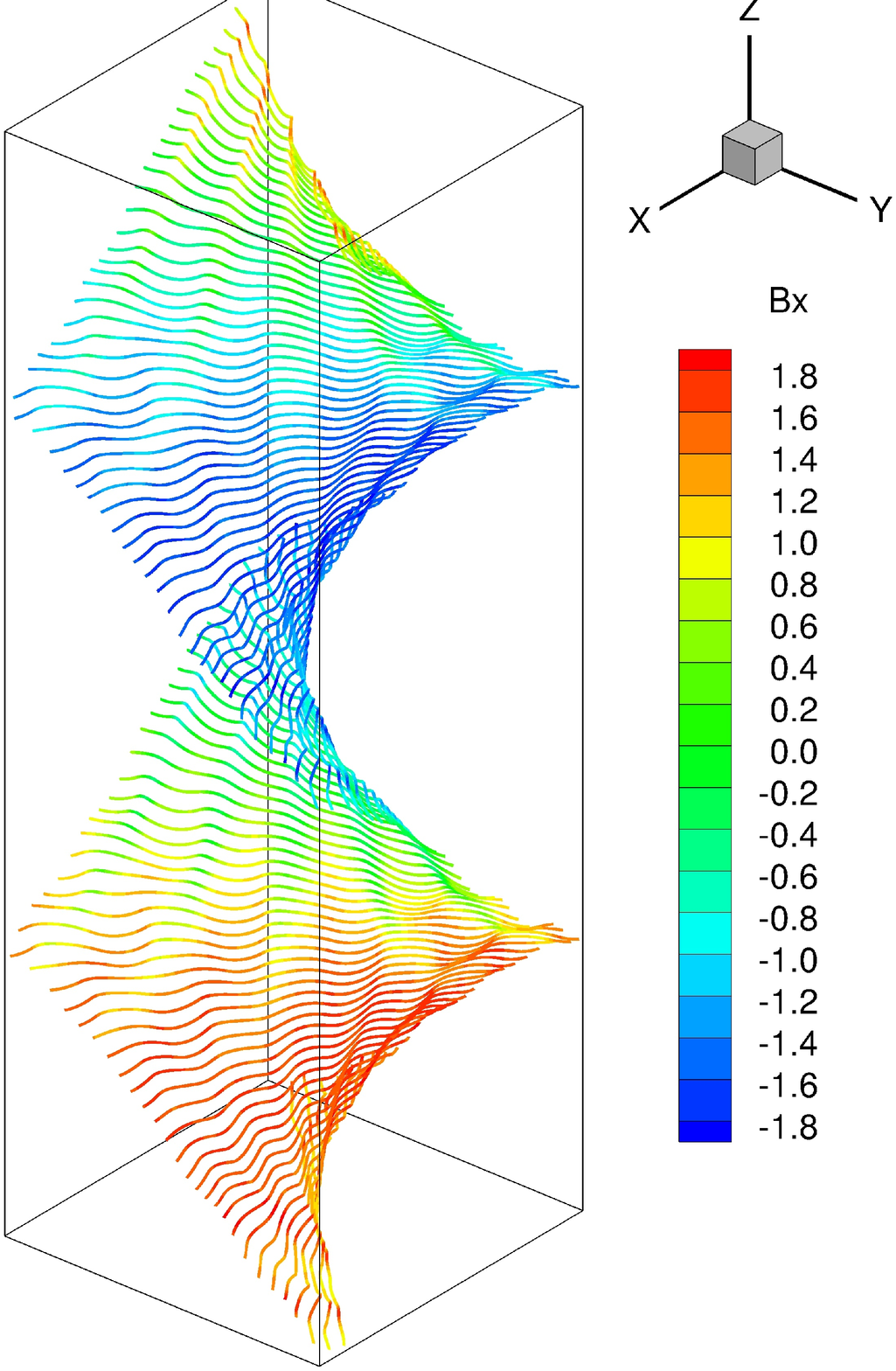}
\caption{Streamlines of the magnetic field slightly above the dynamo
  threshold 
  (left: model A with 16 cells, ${\rm{Rm}}_{\rm{loc}}=5.2$, 
  right: model B with 25 rods, ${\rm{Rm}}_{\rm{loc}}=1.6$)
  The color coding denotes the $x-$ component of the magnetic field. 
  Note the different orientation sense of the helical structure. The
  undulations reflect the small scale field components.\label{fig::streamlines}}
\end{figure}
The typical structure of the magnetic field just above the dynamo
threshold is shown in Figure~\ref{fig::streamlines}. The field
geometry is remarkably similar for both models and essentially
describes a large scale helical pattern dominated by the horizontal
components. The small scale field is visible in terms of little
undulations on top of the large scale structure.

\subsubsection{${\alpha}$-- and ${\beta}$-- effect}

\begin{figure}[h!]
\includegraphics[width=8.0cm]{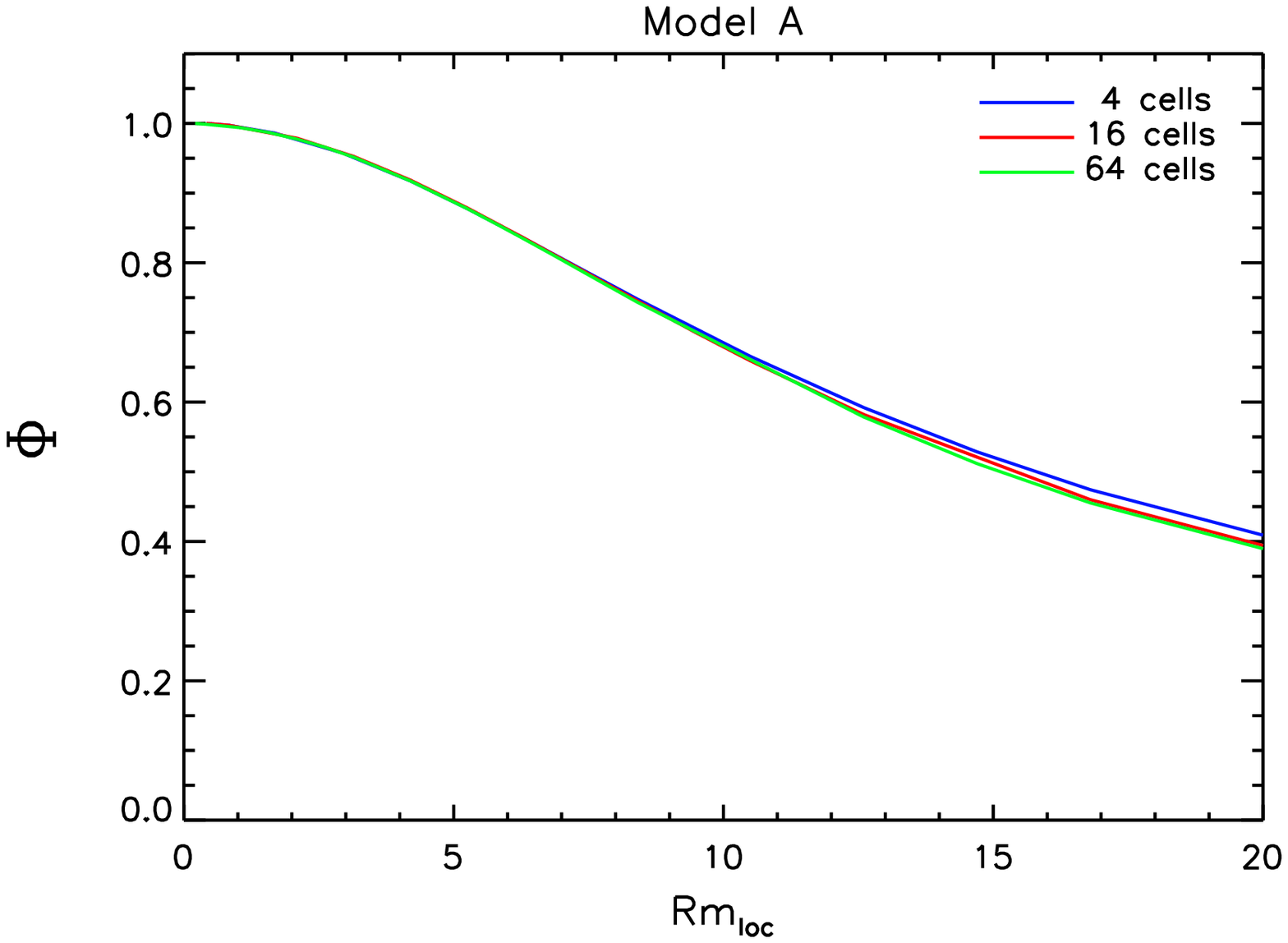}
\includegraphics[width=8.0cm]{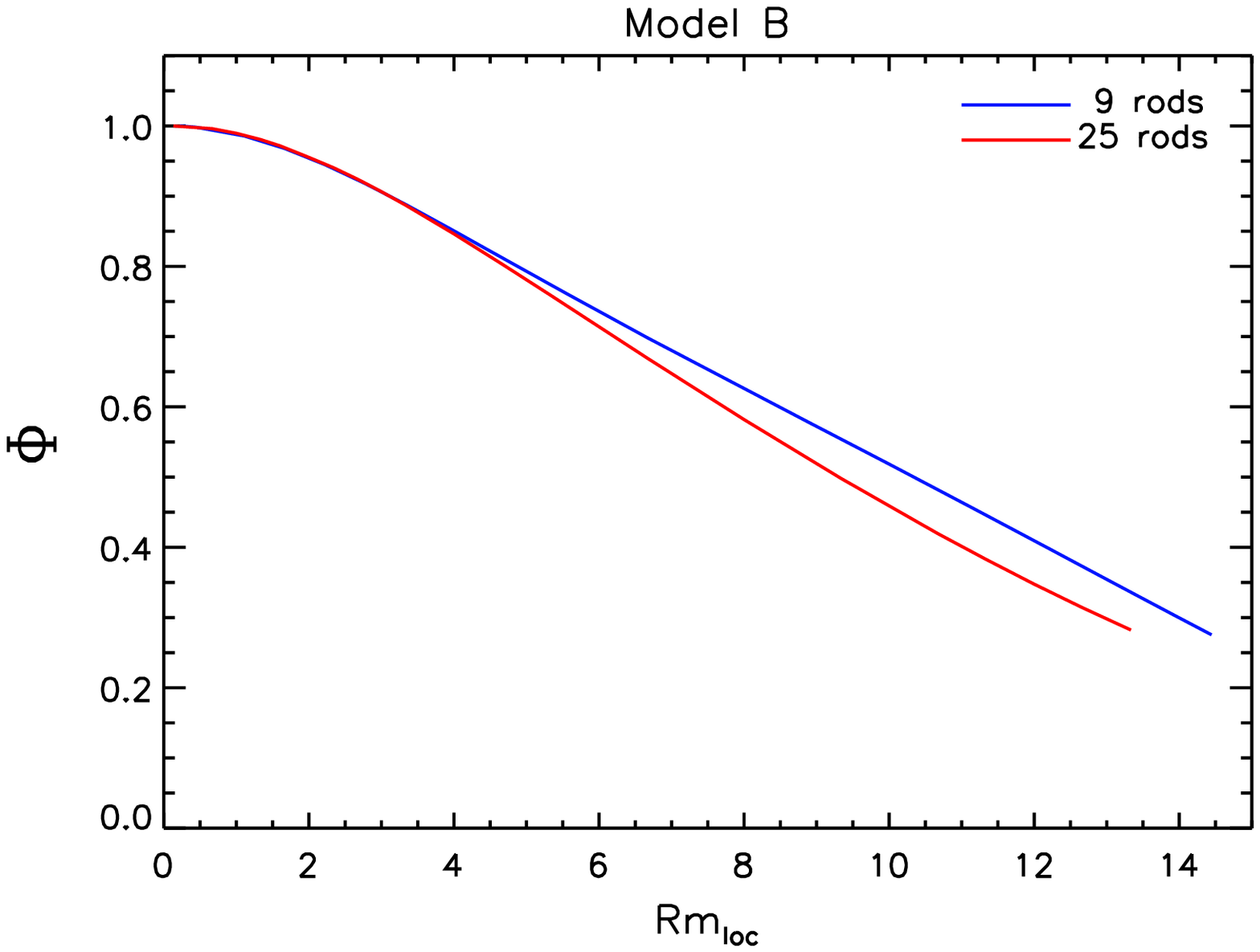}
\caption{$\Phi=\alpha_{\perp}D/(K\eta {\rm{Rm}}_{\rm{loc}}^2)$ versus
  ${\rm{Rm}}_{\rm{loc}}$ for different 
  configurations characterized by an increasing number of helical
  cells. Left: flow model A.
  Right: flow model B.}\label{fig::alpha_vs_rm_mur1}
\end{figure}

We start with a uniform permeability distribution with $\mur=1$ for
both the fluid and the solid internals.  The resulting $\alpha$-effect
is qualitatively in accordance with the results from
\citename{2002NPGeo...9..171R}
\citeyear{2002NPGeo...9..171R,2002MHD....38...41R} for $\alpha$ in
case of an ideal Roberts flow.  Similarly, we write for the
$\alpha$-coefficient
\begin{equation}
\alpha_{\perp}=K\frac{\eta}{D}{\rm{Rm}}_{\rm{loc}}^2\Phi({\rm{Rm}}_{\rm{loc}})\label{eq::alpha}
\end{equation}
with a constant $K$ ($K\approx 0.026$ for model A and $K\approx 0.066$
for model B) and a non-analytic function $\Phi$ that only depends on
${\rm{Rm}}_{\rm{loc}}$.  Figure~\ref{fig::alpha_vs_rm_mur1} shows the
behavior of $\Phi$ versus the flow amplitude ${\rm{Rm}}_{\rm{loc}}$
for three different realizations of model A (with $4, 16$ and $64$
helical cells, left panel) and for two realizations of model B ($9$
and $25$ rods, right panel).
Note, that the normalization factor $K$ is universal for each model
and does not depend on the cell size $D$.  Qualitatively, the behavior
of the function $\Phi$ is similar for both flow models.  $\Phi$
approaches its maximum value for ${\rm{Rm}}_{\rm{loc}}\rightarrow 0$
and decreases monotonically with increasing ${\rm{Rm}}_{\rm{loc}}$, so
that, presumably, $\Phi$ will asymptotically approach zero for very
large flow amplitudes.  Here, we are limited to
${\rm{Rm}}_{\rm{loc}}\la 20$ for model A and to ${\rm{Rm}}_{\rm{loc}}
\la 14$ for model B because above these values the occurrence of small
scale dynamo action with exponentially growing small scale field
prevents a reliable estimation oft the mean-field coefficients. The
onset of small scale dynamo action occurs at smaller $\rml$ for a
larger number of cells, so that the models with the largest $N$
determine the largest achievable $\rml$.  Nevertheless, both models
are already highly overcritical at this $\rm{Rm}_{\rm{loc}}$ so we are
still able to discuss the behavior around the onset of dynamo action
(which is of main interest in the present context).

\begin{figure}[h!]
\includegraphics[width=8.0cm]{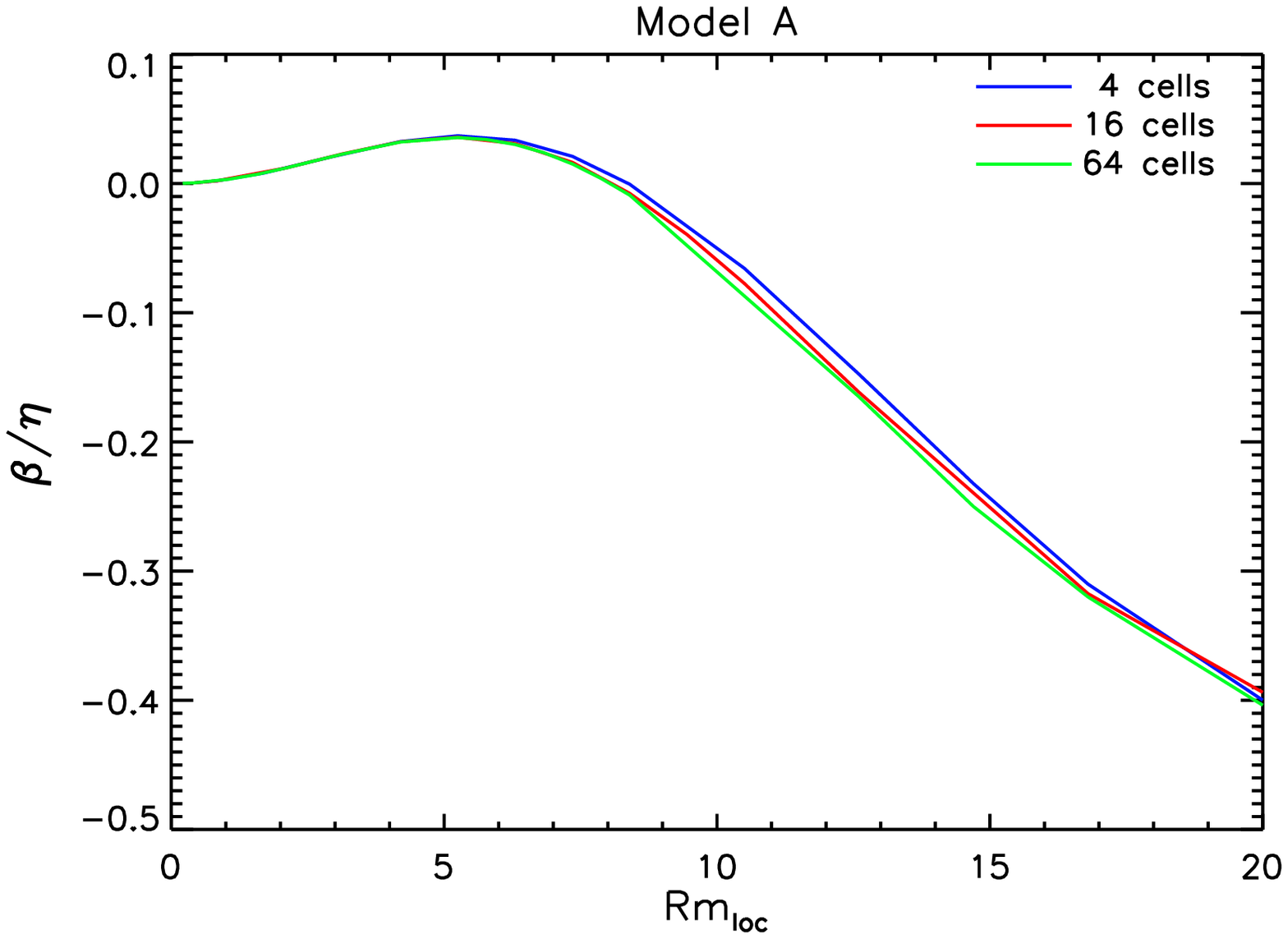}
\includegraphics[width=8.0cm]{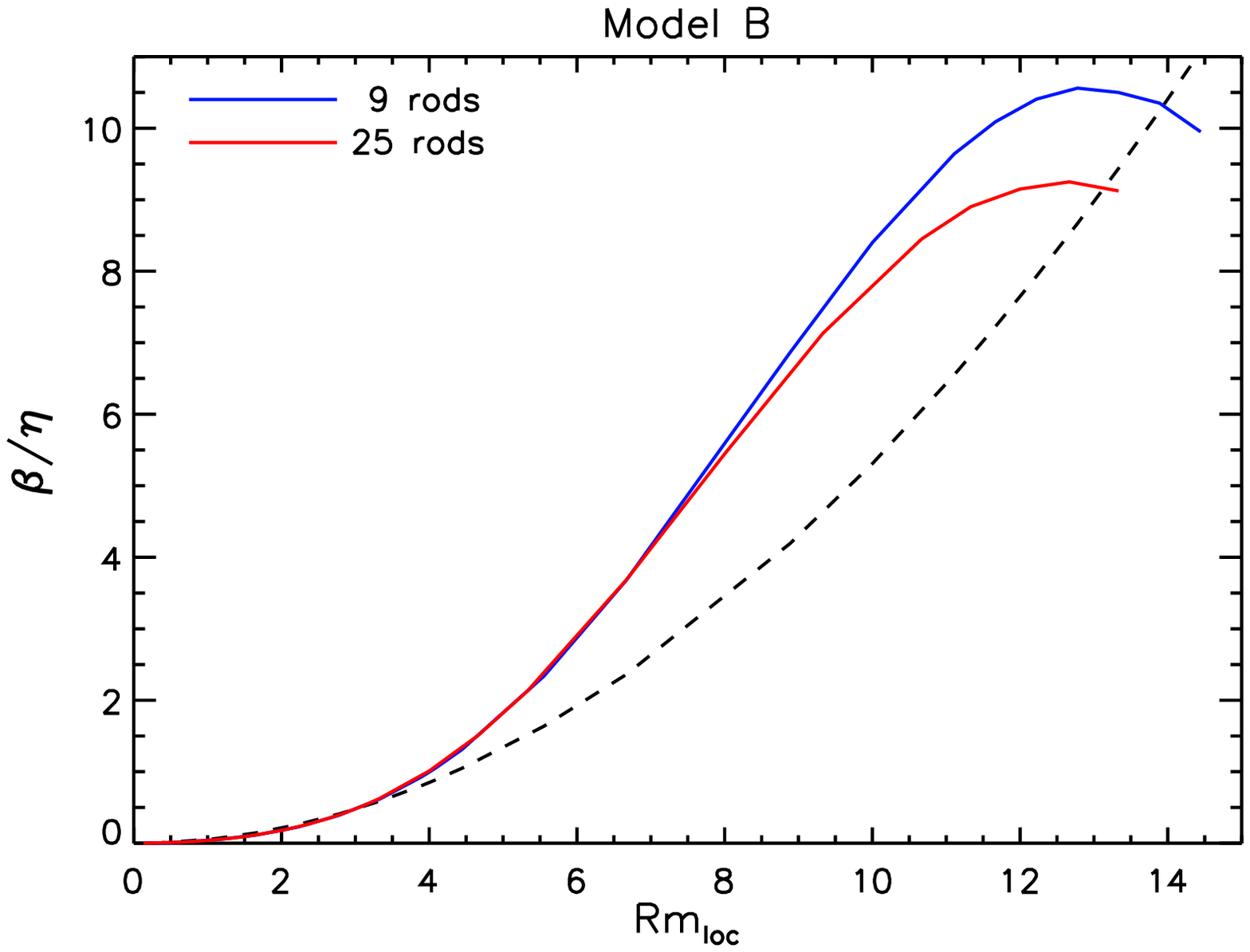}
\\[-3.8cm]
\hspace*{1.4cm}\includegraphics[width=4.0cm]{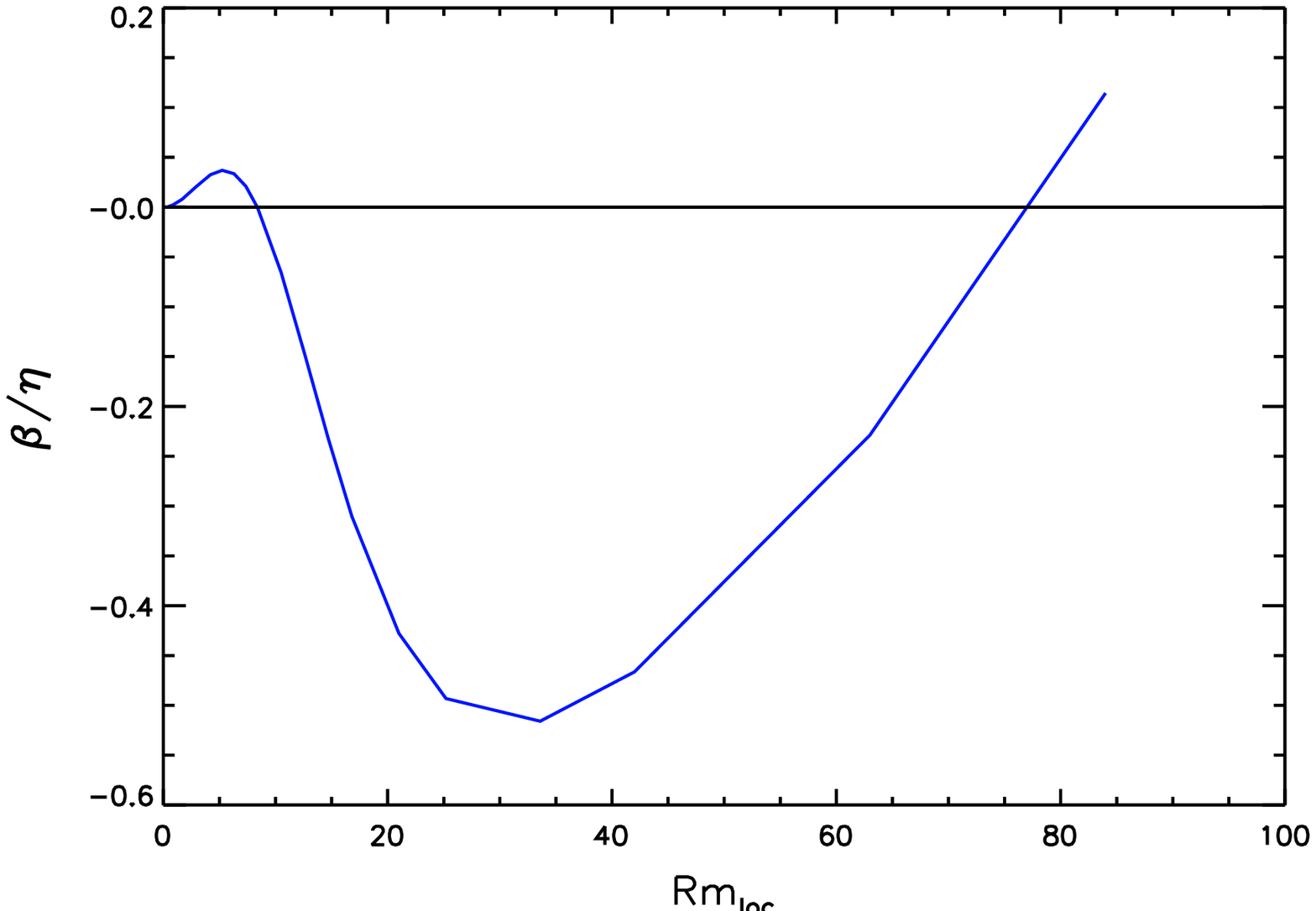}
\\[0.5cm]
\caption{$\beta$ versus ${\rm{Rm}}_{\rm{loc}}$ for different
  configurations of flow model A and B. (left: model A, right: model
  B). The insert plot in the left panel shows
  the behavior of $\beta$ for very large ${\rm{Rm}}_{\rm{loc}}$
  (computed from the model with $4$ helical cells). 
  The dashed black curve in the right panel shows a fitting function $\sim 0.062
  {\rm{Rm}}_{\rm{loc}}^2$ (fitted to the red curve for $\rml \la 3.33$).}\label{fig::beta_vs_rm_mur1}
\end{figure}

Regarding the coefficient $\beta$, we find significant differences
between both models (figure~\ref{fig::beta_vs_rm_mur1}).
To our knowledge, no analytic expressions for $\beta$ beyond the
second order correlation approximation (SOCA), in which
$\vec{u}\times\vec{b}-\overline{\vec{u}\times\vec{b}}=0$ is assumed,
are available in the literature (see, e.g., \citeasnoun{tilgner01} for
an expression for $\beta$ using SOCA). The restricted validity of SOCA
is shown in \citeasnoun{2008A&A...482..739B} where mean-field
coefficients obtained from the test-field method with and without SOCA
are compared. Since in our model the
preconditions for SOCA are not met, we refrain from a similar
analysis.
Surprisingly, we do not observe any dependency on the cell size for
model A and only a weak dependence in case of model B. Thus, the
$\beta$-effect mainly depends on $\rml$ and is independent of the
characteristic wave number of the small scale flow (at least within
the rather restricted range of flow scales that has been examined for
this study).
The most striking property of the $\beta$-coefficient in model A is
the transition to negative values around ${\rm{Rm}}_{\rm{loc}}\approx
8$\footnote[7]{A similar behavior has already been observed in certain
  parameter regimes for the Roberts flow examined in
  \citename{2008A&A...482..739B} \citeyear{2008A&A...482..739B}.}.  In
general, the $\beta$-effect is associated with an enhancement of the
magnetic diffusivity due to the small scale motion and hence should be
positive.  The occurrence of a negative $\beta$ effect can be
explained by the presence of two contributions related to field
gradients in the $z$-direction that cannot be separated from each
other in our configuration: the anisotropic part of the magnetic
diffusivity $\beta_{\perp}$ (which is assumed to be positive) and the
term related to the symmetric part of the field gradient tensor
described by $\tilde{\beta}$ in equation (\ref{eq::simp_emf}). For the
second contribution no restrictions for the sign are known so that the
sum of both terms can become negative.  A negative $\beta$ may be
helpful for dynamo action, but in our models the sum of $\beta$ and
the ``normal'' diffusivity $\eta$ (which is set to unity in all runs)
always remains positive (see insert plot in the left panel of
figure~\ref{fig::beta_vs_rm_mur1}) so that the consideration of
$\beta$ results ``only'' in a reduction of the overall diffusivity
$\eta_{\rm{tot}}=\eta+\beta$ (if we neglect that $\beta$ is related to
an anisotropic contribution).

The relative amplitude of $\beta$ is much larger in model B with
$\beta$ exceeding $\eta$ by up to a factor of $10$, and we do not find
a negative $\beta$-effect within the achievable parameter regime.
However, a local maximum of $\beta$ exists around $\rml\approx 13$ and
it cannot be ruled out that the further development of $\beta$ follows
a similar path as in model A but for larger values of $\rml$ and
$\beta$.

For small $\rml$ the behavior of $\beta$ is roughly 
proportional to ${\rm{Rm}}_{\rm{loc}}^2$ (see the black dashed curve
in the right panel in figure~\ref{fig::beta_vs_rm_mur1}) which is in
accordance with measurements of the $\beta$-effect 
in the Perm experiment \cite{2010PhRvL.105r4502F,2012PhRvE..85a6303N}.

\subsubsection{Comparison between fully resolved models and mean-field models}

In the following, the $\alpha$- and $\beta$-coefficients presented in
figure \ref{fig::alpha_vs_rm_mur1} and \ref{fig::beta_vs_rm_mur1} will
be used as an input for mean-field dynamo simulations.  The
corresponding equation includes the mean flow $\overline{\vec{U}}$
obtained from horizontal averaging of equations (\ref{eq::flowA}) or
(\ref{eq::flowB}), the EMF given by equation~(\ref{eq::simp_emf}), and
a diffusive term $\propto\nabla\times\overline{\vec{B}}$ that involves
an effective (mean) diffusivity $\overline{\eta}$.  The mean
diffusivity $\overline{\eta}$ is computed by dividing the ``normal''
(uniform) diffusivity $\eta$ by the horizontal average of
$\mur(\vec{r})$:
\begin{equation}
\overline{\eta}=\frac{\eta}{L^{-2}\int\limits_{x,y}(\mur(\vec{r}))dxdy}.
\end{equation}
with $L$ the horizontal width of the computational domain.
The resulting mean-field induction equation reads: 
\begin{flalign}
\frac{\partial\overline{\vec{B}}}{\partial t}&=
\nabla\!\times\!\left(\!\overline{\vec{U}}\!\times\!\overline{\vec{B}}\!-\!\alpha_{\perp}(\overline{\vec{B}}
-(\hat{\vec{z}}\!\cdot\!\overline{\vec{B}})\hat{\vec{z}})
-\gamma\!\times\!\overline{\vec{B}}
-\overline{\eta}\nabla\!\times\!\overline{\vec{B}}
\!-\!\beta\hat{\vec{z}}\!\times\!\frac{d\overline{\vec{B}}}{dz}\!-\!\delta_2(\hat{\vec{z}}\!\cdot\!\nabla)\overline{\vec{B}}\!\right),&
\label{eq::mf_simple}
\end{flalign} 
where we additionally specified the terms related to $\gamma$ and
$\delta_2$ which mostly have no influence on the growth-rates.
For $\gamma$ this is true for all runs, whereas for $\delta_2$ we
assume a beneficial impact for dynamo action in model B in case of
large $\mur$ and large $\rml$ (see below).

Figure \ref{fig::gr_vs_rmperp} shows
the growth-rates obtained from the fully resolved models (FRM)
that have been used to compute the mean-field coefficients in the
previous section (solid curves) in comparison with the 
growth-rates obtained from the mean-field models (MFM) at various
${\rm{Rm}}_{\rm{loc}}$ (dashed curves and stars). 
\begin{figure}[h!]
\includegraphics[width=8.0cm]{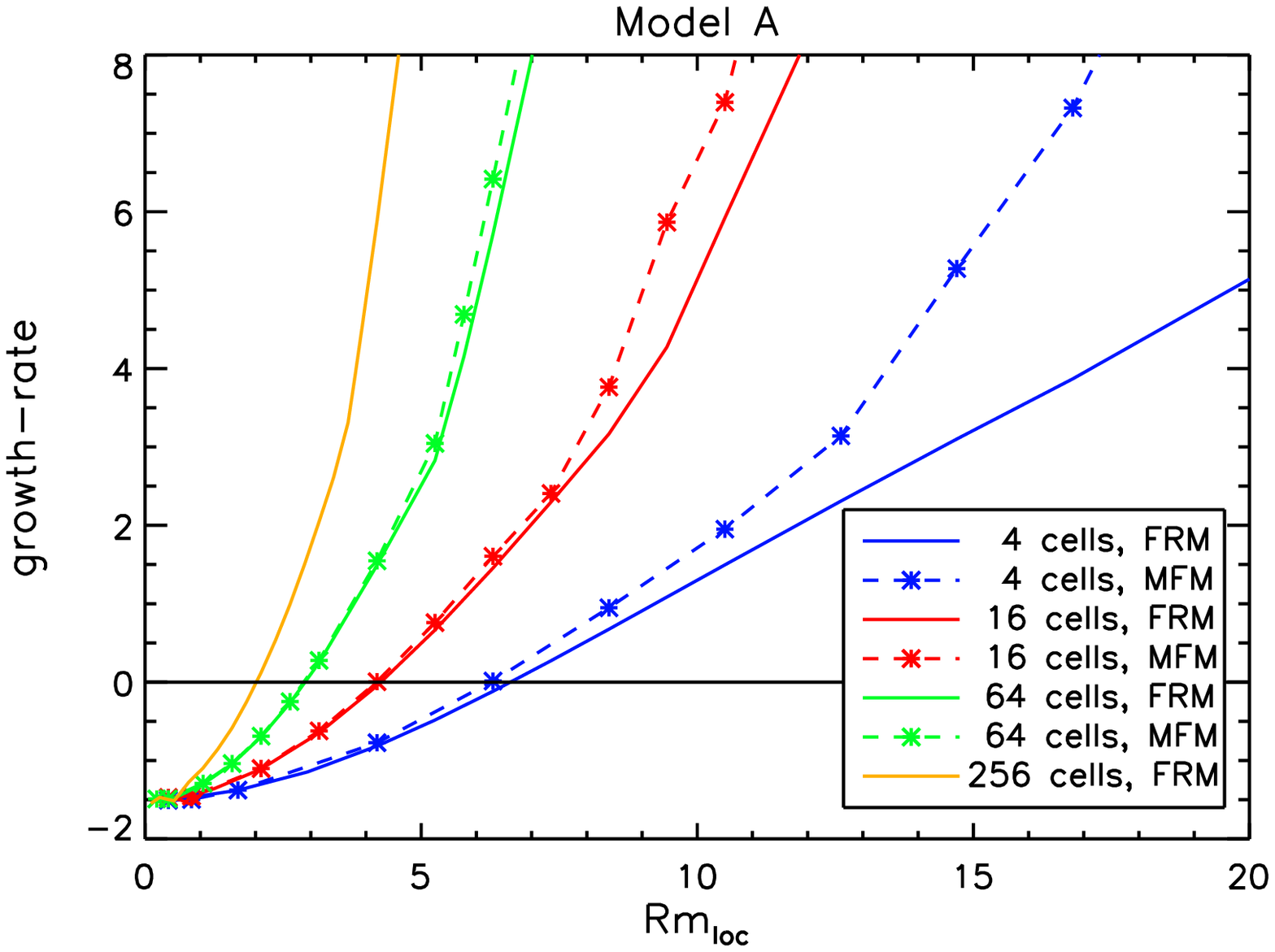}
\includegraphics[width=8.0cm]{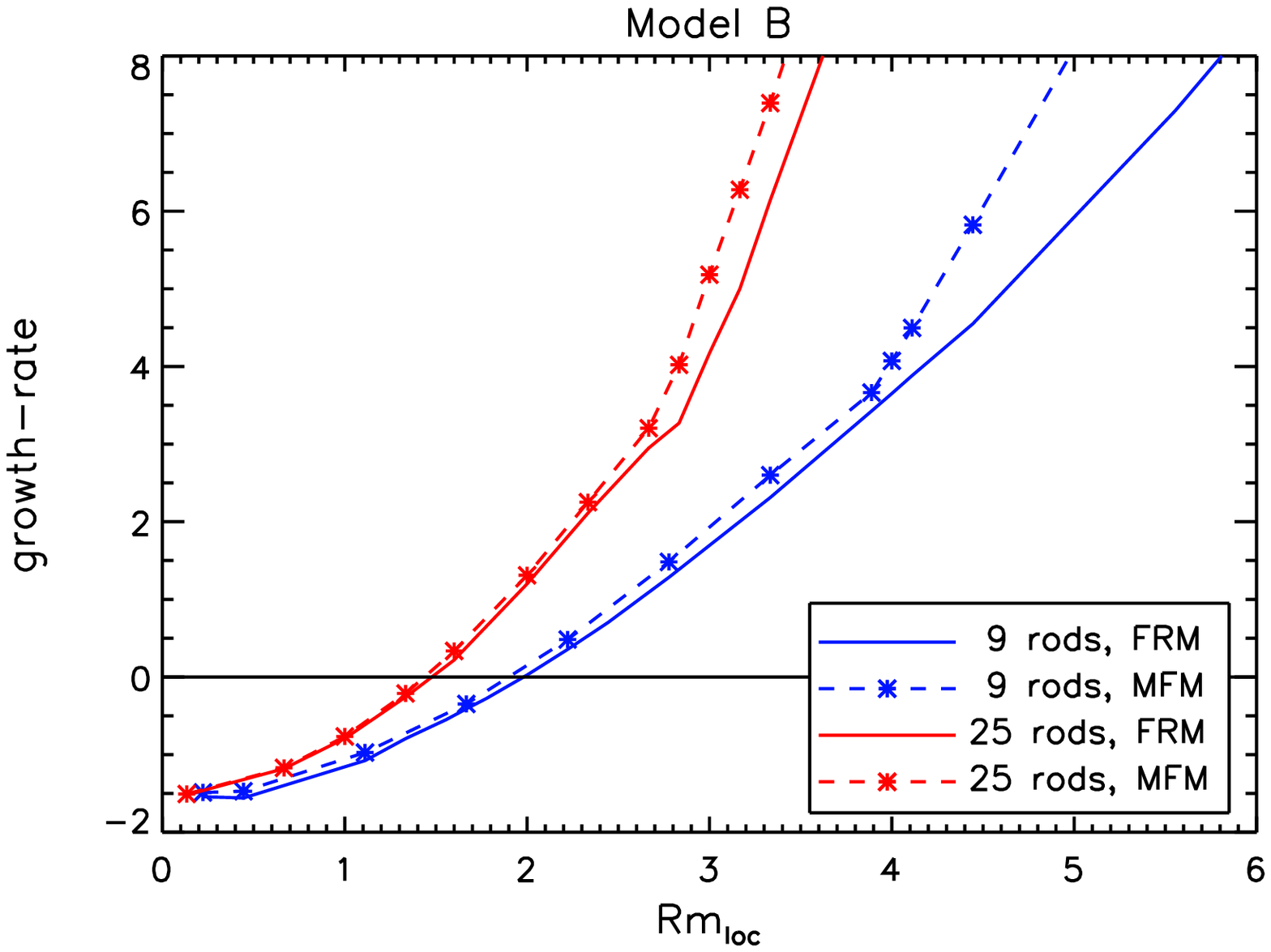}
\caption{Growth-rate versus ${\rm{Rm}}_{\rm{loc}}$ for
  different configurations of model A (left panel) with $4, 16, 64,
  \mbox{ and } 256$ cells and model B (right panel) with 9 and 25 rods, respectively. 
  The solid curves denote the growth-rates from the fully resolved
  models (FRM) and 
  the dashed curves denote the growth-rates for mean
  field models (MFM) that have been computed using the mean-field
  coefficients obtained for the specific ${\rm{Rm}}_{\rm{loc}}$
  that are marked with the stars.
}\label{fig::gr_vs_rmperp}
\end{figure}
We obtain quite a good agreement between FRM and MFM if the system is
not strongly overcritical.  The agreement becomes better for an
increasing number of helical eddies, i.e., for an increasing scale
separation which provides a better fulfilment of the prerequisites for
applying the mean-field theory. 
The rather large deviations in the strongly
overcritical regime can be explained by a transition of the vertical
wavenumber of the leading eigenmode from $n=1$ to $n=2$. The higher wavenumber
is not incorporated by the particular
vertical wave number of the applied testfields which are $\propto
\cos(z)$ and $\propto\sin(z)$.  In principle this issue could
be attacked by computing mean-field coefficients for testfields
$\propto \cos(n z)$ and $\propto \sin(n z)$ with $n=2,3,4,...$
and including these contributions in the mean field models
\citeaffixed{2008A&A...482..739B}{see, e.g.,}.

\begin{figure}[h!]
\includegraphics[width=8.0cm]{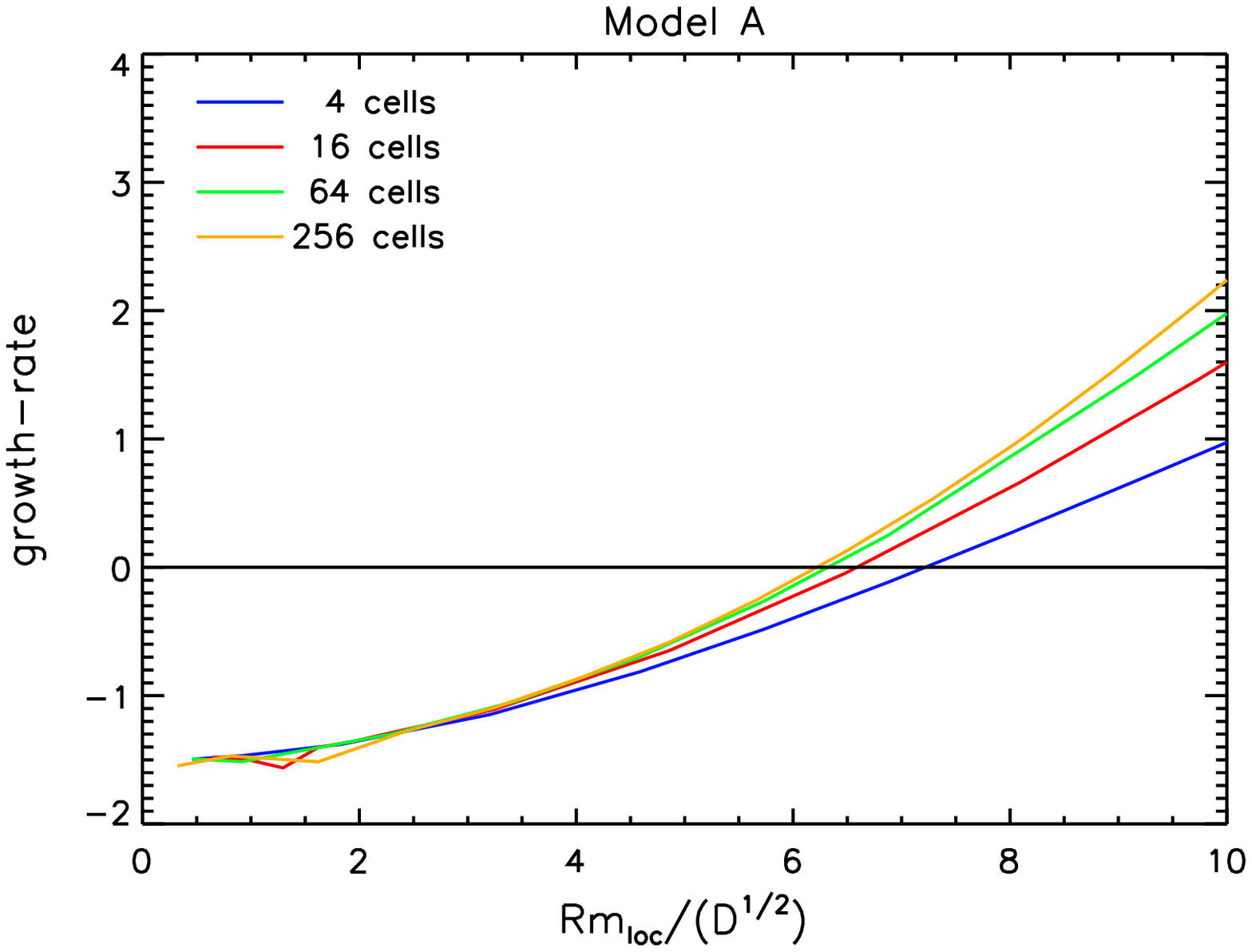}
\includegraphics[width=8.0cm]{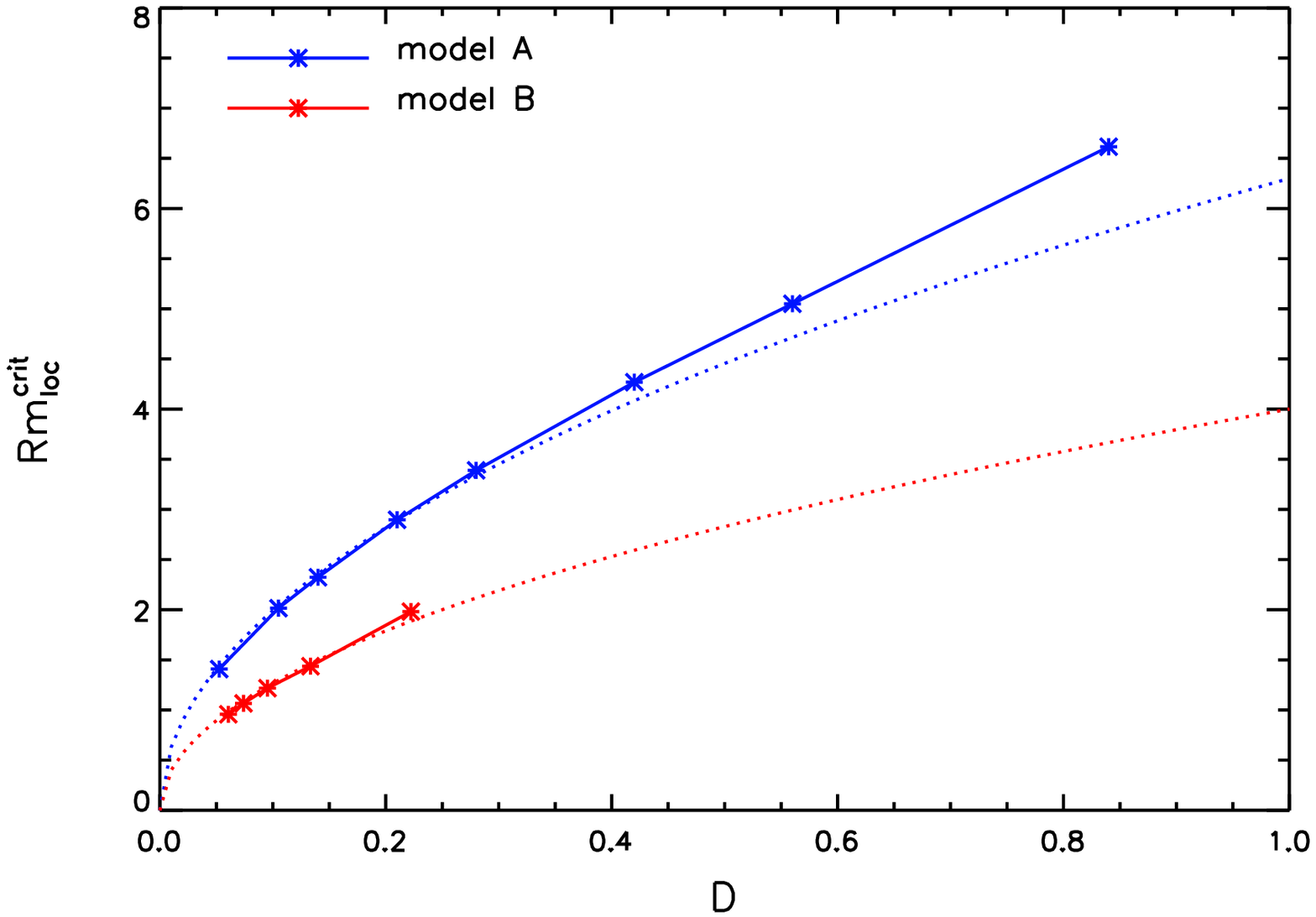}
\caption{Left panel: growth-rate versus ${\rm{Rm}}_{\rm{loc}}$ divided by
  $\sqrt{D}$ for model A.
  Right panel: Critical value of the local magnetic Reynolds number in
  dependence of the cell size $D$. The dotted curves show a fit
  proportional to $\sqrt{D}$.}\label{fig::gr_vs_rm_modelA_scaled}
\end{figure}

The growth-rates presented in figure~\ref{fig::gr_vs_rmperp} show that
a reduction of the scale of a single helical cell (or an increase of
the number of helical cells) improves the dynamo properties of the
system.  The increase in growth-rate with decreasing $D$ follows a
typical scaling law, which becomes apparent from the left panel of
figure~\ref{fig::gr_vs_rm_modelA_scaled} where the growth-rates are
plotted against $\rml$ divided by $\sqrt{D}$.  The scaling is almost
perfect for small magnetic Reynolds numbers and convergence arises
when changing the flow pattern from 64 to 256 cells (compare green and
orange curve in the left panel in
figure~\ref{fig::gr_vs_rm_modelA_scaled}).  A similar scaling is
obtained for the critical magnetic Reynolds number
${\rm{Rm}_{\rm{loc}}^{\rm{crit}}}$ that is required for the onset of
dynamo action.
The behavior of ${\rm{Rm}_{\rm{loc}}^{\rm{crit}}}$ for decreasing cell
size $D$ can be derived assuming that the onset of dynamo action is
governed by some global magnetic Reynolds number. This quantity may be
defined on the basis of an effective length scale that is given by the
linear number of cells (which in our quadratic configuration is equal
to $\sqrt{N}$) multiplied with the typical scale of a single cell $D$.
Then the onset of mean-field dynamo action is determined by
${\rm{Rm}}_{\rm{glob}}^{\rm{crit}}\sim \sqrt{N}
D\alpha^{\rm{crit}}/\eta$.  For a large number of cells (corresponding
to a small $D$) we have $\rml\ll 1$, so that using
equation~(\ref{eq::alpha}) with $\varPhi\rightarrow 1$, we can write
$\sqrt{N}D\alpha^{\rm{crit}}/\eta=\sqrt{N}K({\rm{Rm}}_{\rm{loc}}^{\rm{crit}})^2$. 
Given that we used $\sqrt{N}D=\mbox{const}$, this immediately yields
${\rm{Rm}}_{\rm{loc}}^{\rm{crit}}\sim\sqrt{D}$ which is indeed
confirmed by our results (right panel in
figure~\ref{fig::gr_vs_rm_modelA_scaled}).
A more detailed analysis of the critical magnetic Reynolds number for
a mean field model of the Roberts flow that includes the dependence on
the vertical extension can be found in \citeasnoun{tilgner02}.

\subsection{Walls and rods with ${\mur > 1}$}

In the following, we only examine systems with $16$ eddies
(model A) and $9$ rods (model B) because the consideration of a
permeability distribution with $\mur>1$ extends the necessary
simulation time due to the decreased effective diffusivity, so that we
are limited to smaller systems with lower grid resolution.
\begin{figure}[t!]
\includegraphics[width=7.125cm]{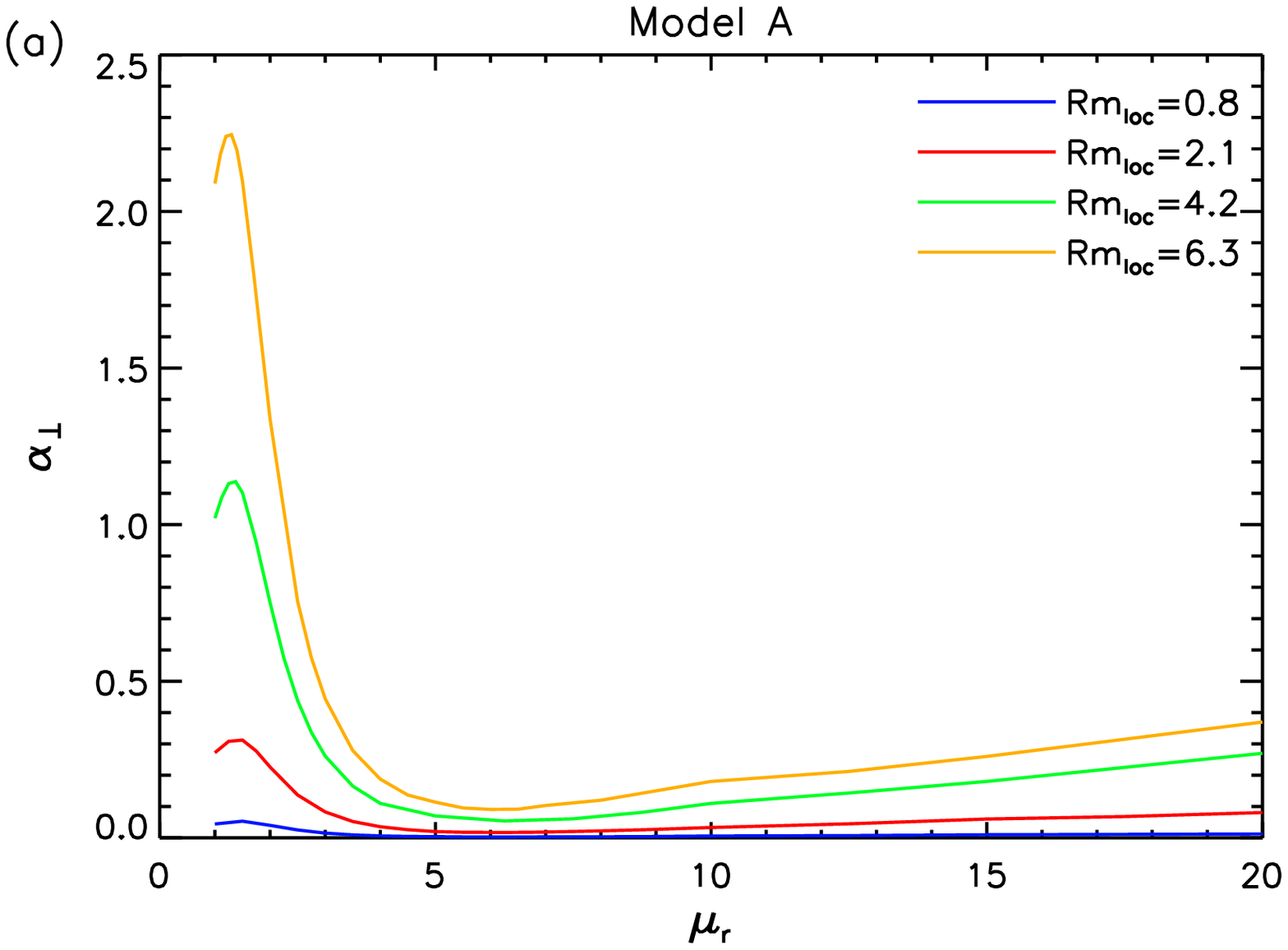}
\nolinebreak[4!]
\includegraphics[width=7.125cm]{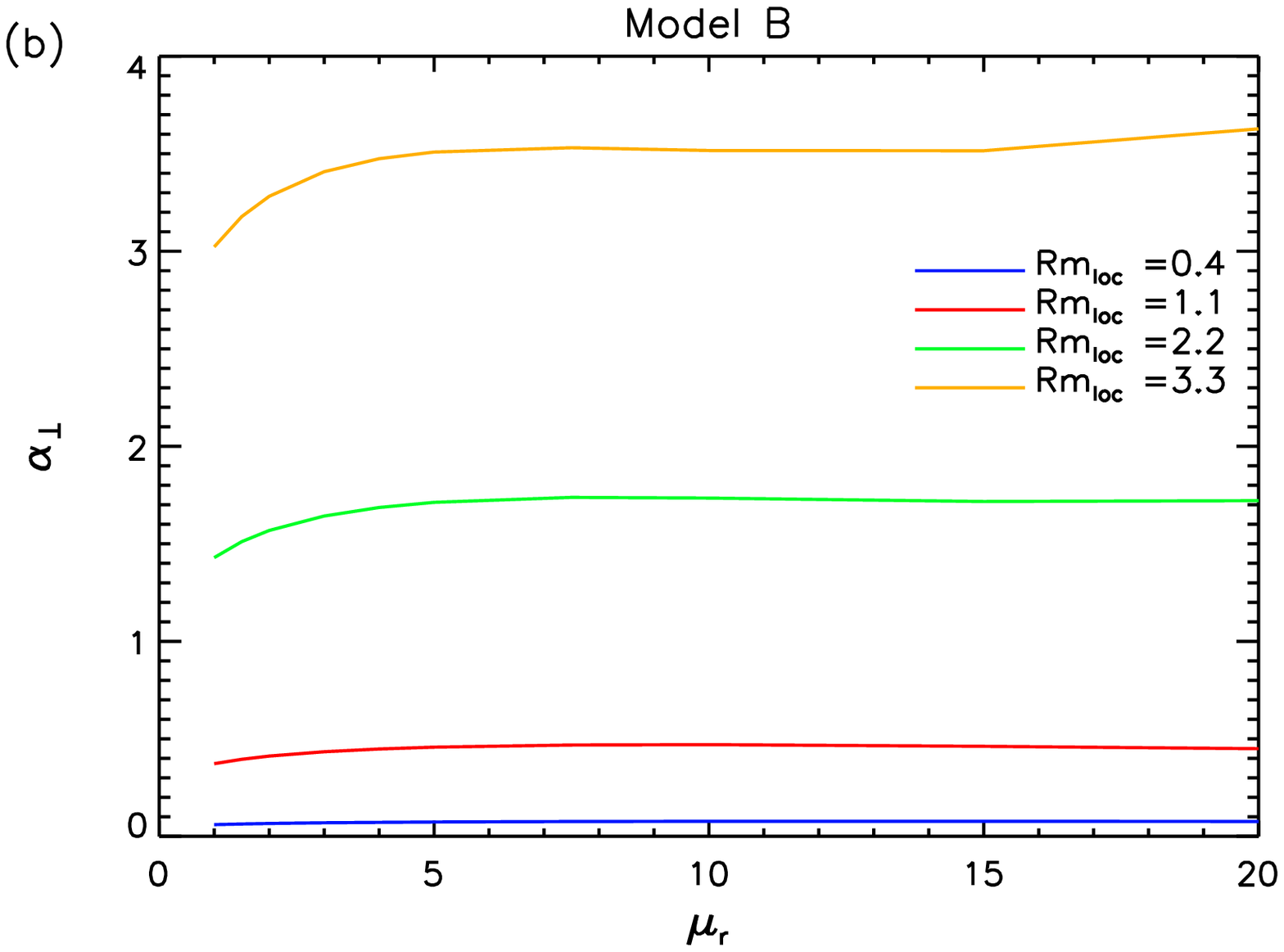}
\\
\includegraphics[width=7.125cm]{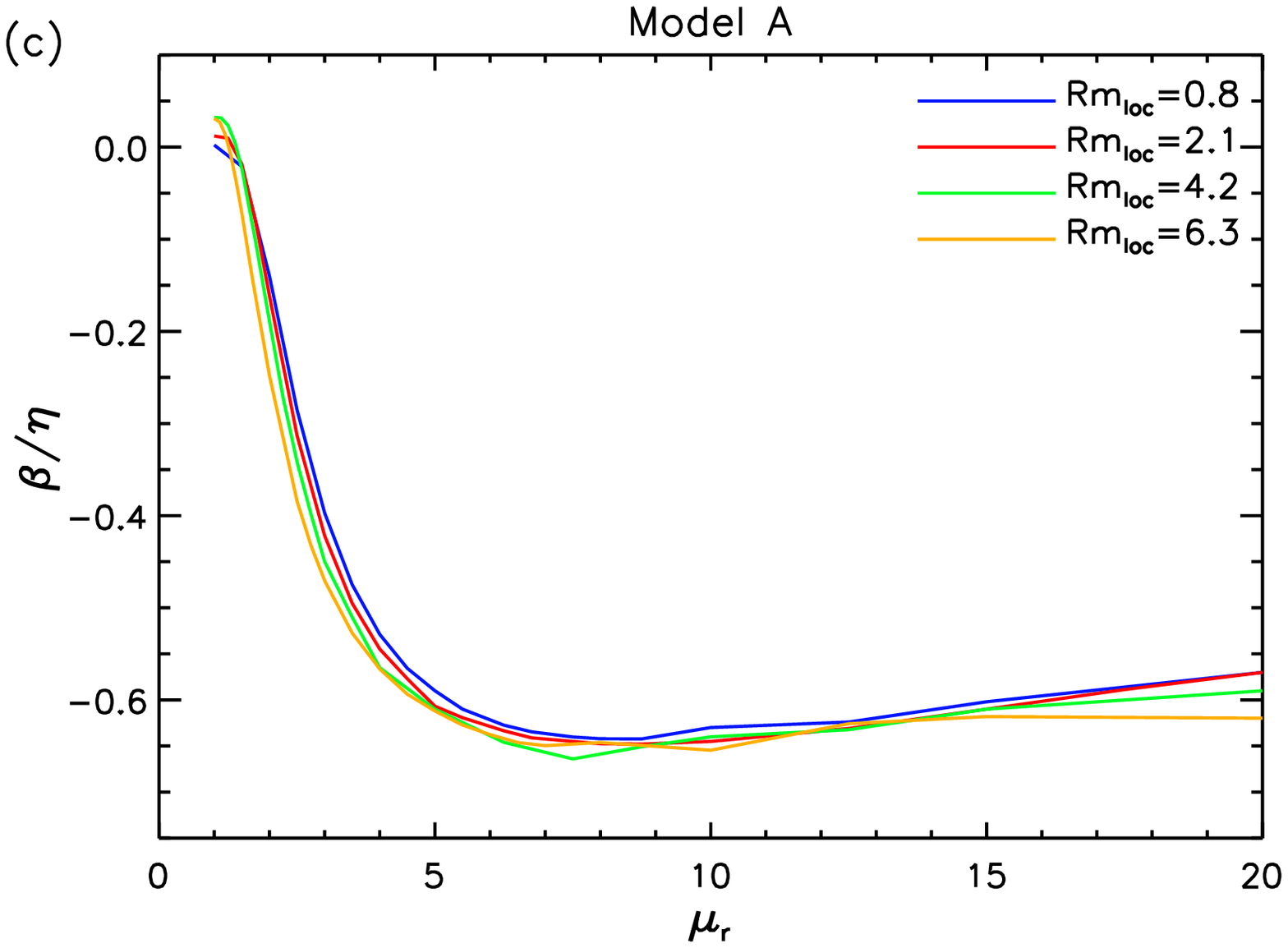}
\nolinebreak[4!]
\includegraphics[width=7.125cm]{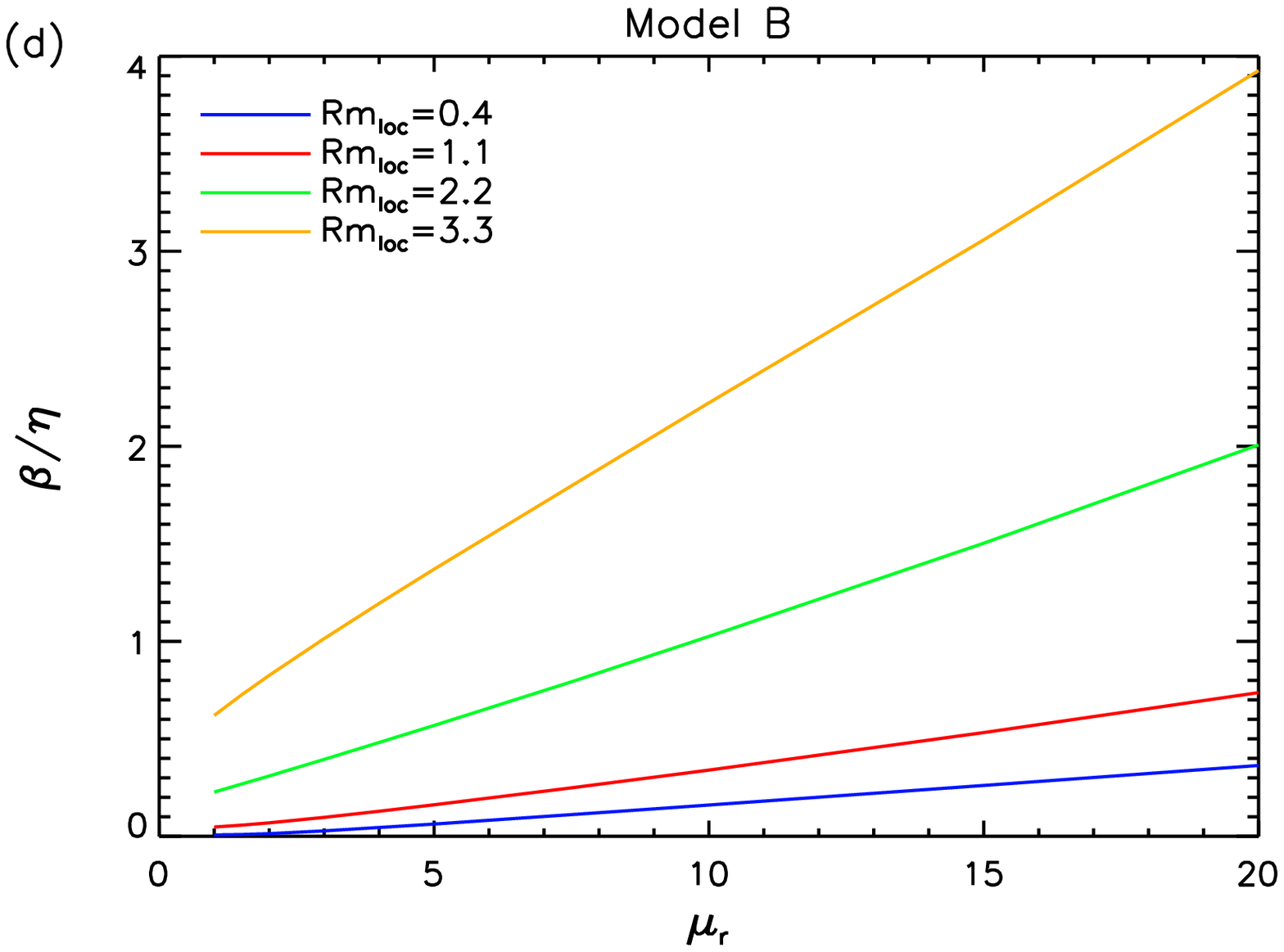}
\\
\includegraphics[width=7.125cm]{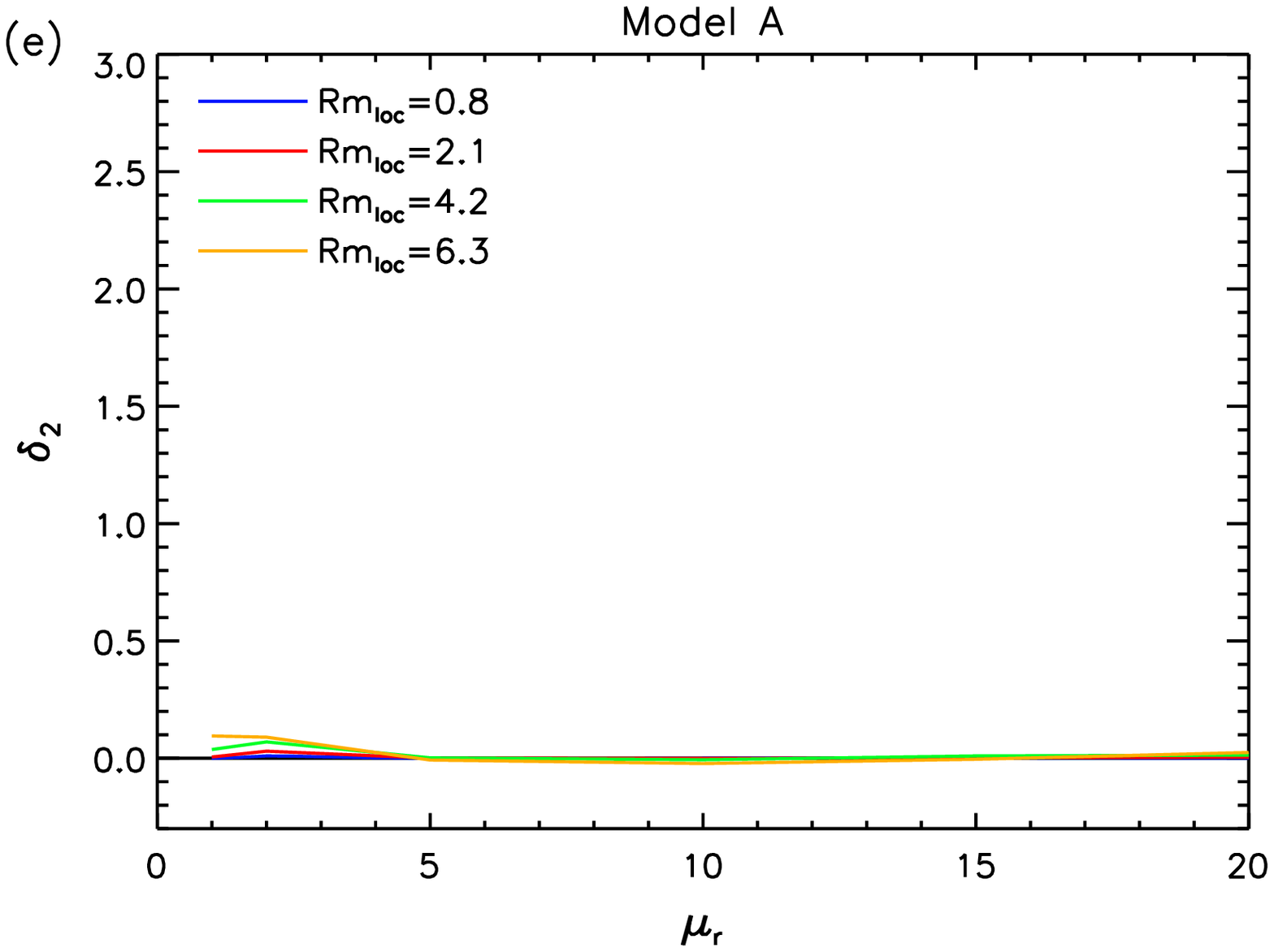}
\nolinebreak[4!]
\includegraphics[width=7.125cm]{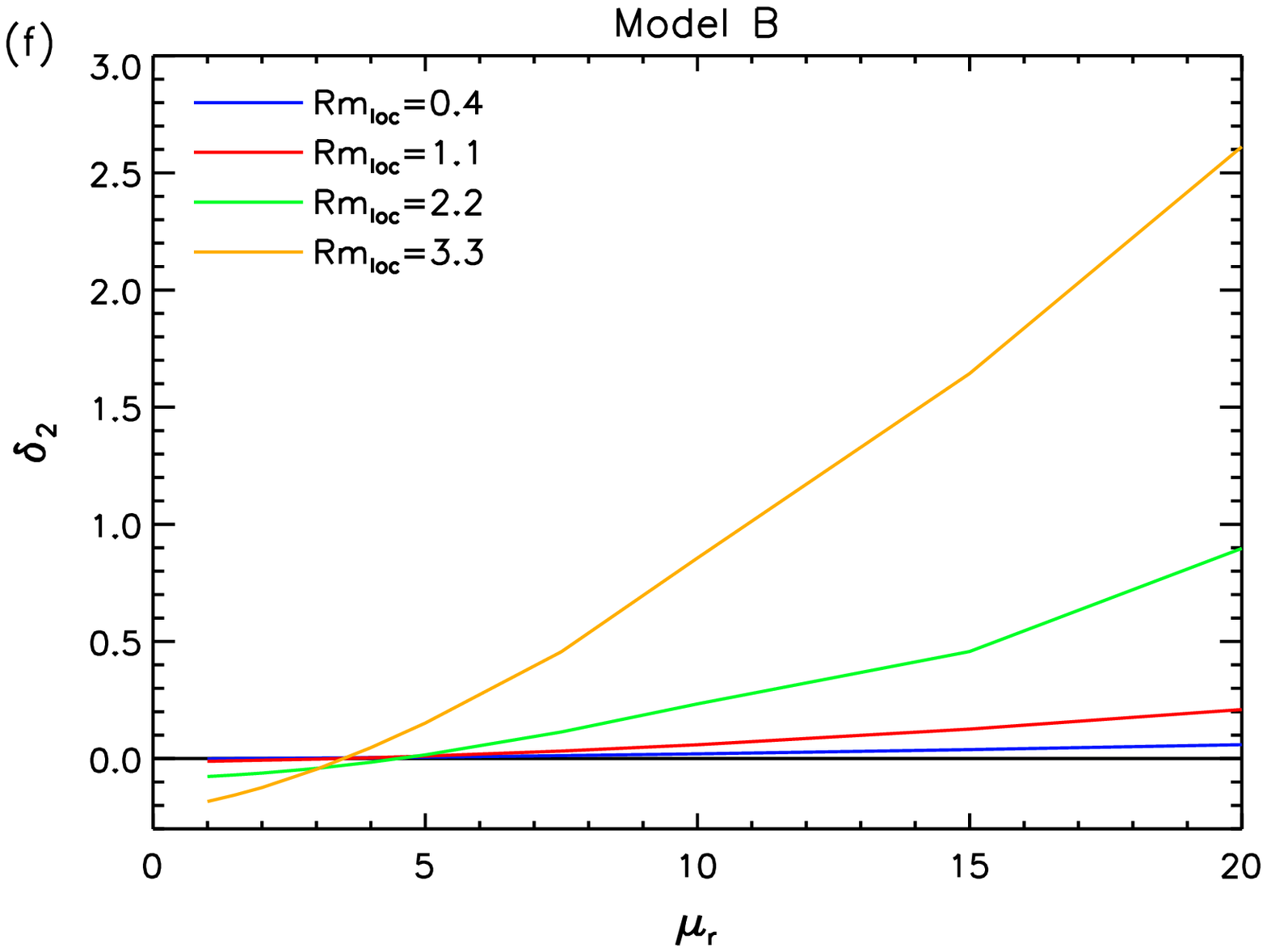}
\\
\includegraphics[width=7.125cm]{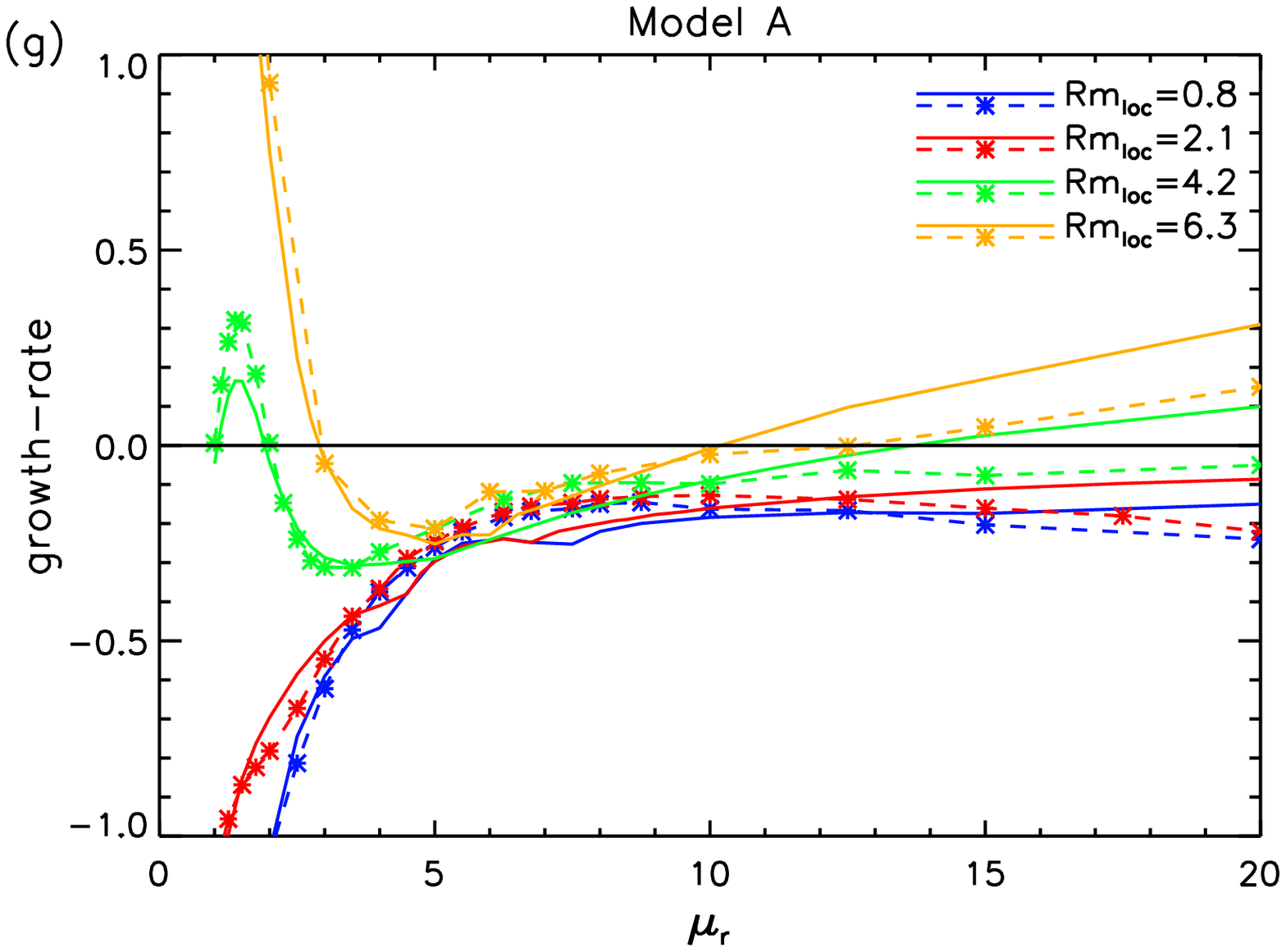}
\nolinebreak[4!]
\includegraphics[width=7.125cm]{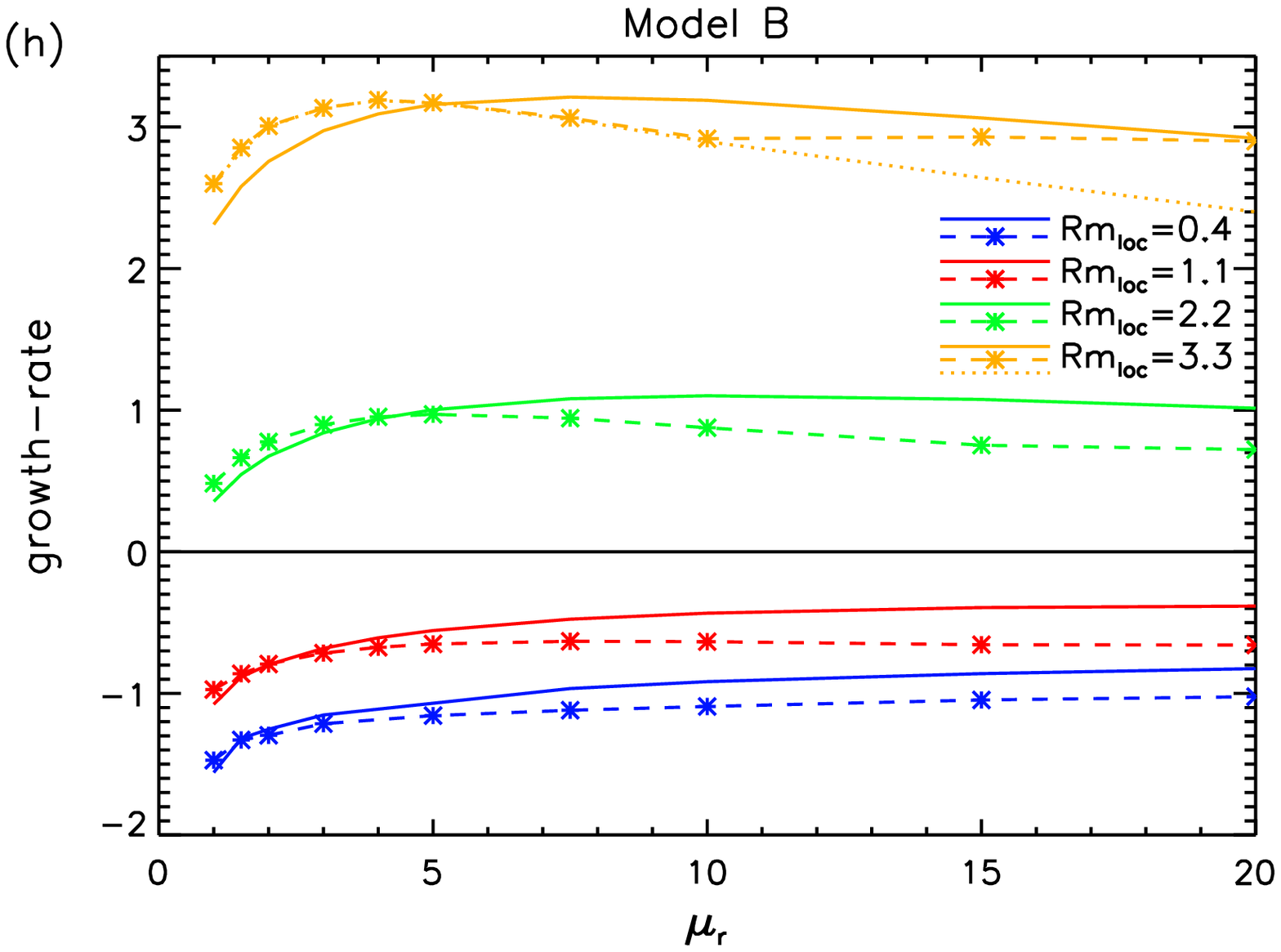}
\caption{
Mean-field coefficients and growth-rates versus $\mur$ for various
$\rml$ (left: model A, right: model B).
Top row: $\alpha_{\perp}$ versus $\mur$,
second row: $\beta$ versus $\mur$,
third row: $\delta_2$ versus $\mur$,
fourth row: growth-rate versus $\mur$ and comparison between FRM
(solid curves) and MFM (dashed curves and stars). The dotted orange
curve in panel (h) shows the MFM growth-rates without $\delta_2$ term.}
\label{fig::mf_gr_vs_mur}
\end{figure}
Not surprisingly, the results become more complex when $\mur > 1$.
Figure~\ref{fig::mf_gr_vs_mur}a and b show the behavior of
$\alpha_{\perp}$ versus $\mur$ for different values of $\rml$.  Here
we refrain from any scaling for $\alpha_{\perp}$ in order to carve out
the direct influence of $\rml$ and/or $\mur$ on $\alpha_{\perp}$.  For
a fixed $\mur$, we always find that $\alpha_{\perp}$ grows with
increasing $\rml$.  However, we find significant differences between
both flow models regarding the dependence on the permeability.  For
model A we observe a significant suppression of $\alpha_{\perp}$ for
small permeabilities (say $\mur<10$), followed by a slow recovery for
further increasing $\mur$.  In contrast, for model B we see a moderate
increase of $\alpha_{\perp}$ for small $\mur$ followed by a saturation
regime for $\mur\ga 5$ in which $\alpha_{\perp}$ becomes largely
independent of $\mur$.

For $\mur$ only slightly above $1$ in model A, we find a sharp maximum
for $\alpha_{\perp}$ around $\mur\approx 1.5$. This maximum is
retrieved again in the corresponding growth-rates (as the
$\beta$-effect has no equivalent peak or drop) but the influence on
the critical magnetic Reynolds numbers remains small (see below).

A significant difference between the two models is also found in the
behavior of the $\beta$-effect (figure~\ref{fig::mf_gr_vs_mur}c
and~\ref{fig::mf_gr_vs_mur}d).  For model A, we see an abrupt
transition to negative values between $\mur=1$ and $\mur\approx 5$,
and $\beta$ remains nearly constant ($\beta\approx -0.6$) for $\mur\ga
5$.  In model B the behavior of $\beta$ is surprisingly simple and
monotonic.  $\beta$ just increases linearly with increasing $\mur$
with the slope increasing according to $\rml$. In particular, we do
not find any indications for a transition to negative values of
$\beta$ with this flow configuration.

In figure~\ref{fig::mf_gr_vs_mur}e and f we additionally present the
behavior of the coefficient $\delta_2$. This coefficient does not play
any role for model A, but $\delta_2$ becomes large in model B for
large $\rml$ and $\mur$. In this parameter regime, we see an
$\alpha$-effect which is independent of $\mur$ whereas $\beta$ is
linearly increasing. This feature would be inconsistent with the
growth-rates, which are also nearly independent of $\mur$, so it
requires an additional term that compensates for the losses from the
$\beta$-effect.  The only possibility in our models stems from the
effects described by $\delta_2$. Indeed, this is confirmed in comparative
mean-field models without the term $\propto\delta_2$ in which we find
a decreasing growth-rate in the limit of large $\rml$ and large $\mur$
(see dotted orange curve in figure~\ref{fig::mf_gr_vs_mur}h).

\begin{figure}[h!]
\includegraphics[width=8cm]{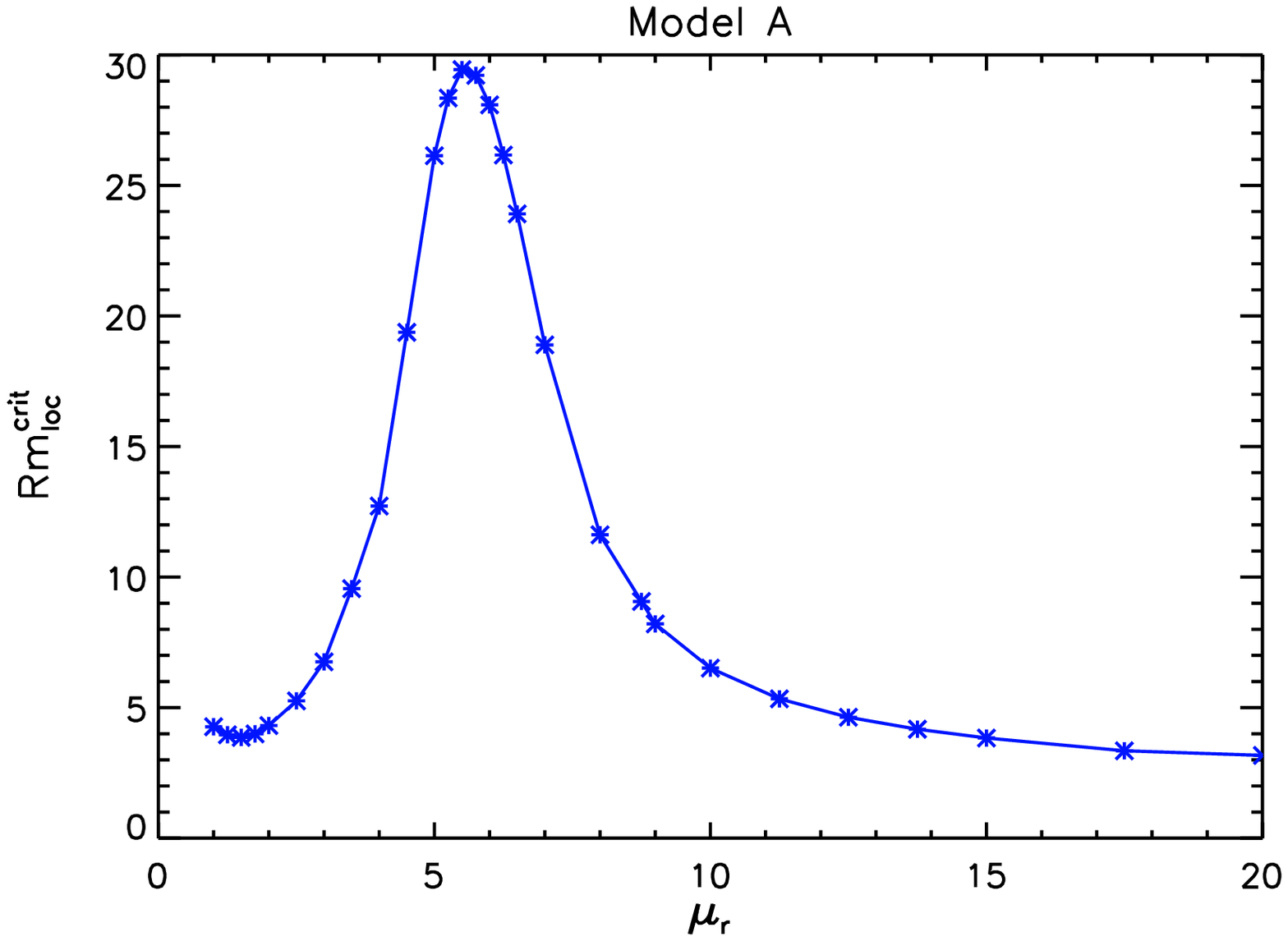}
\includegraphics[width=8cm]{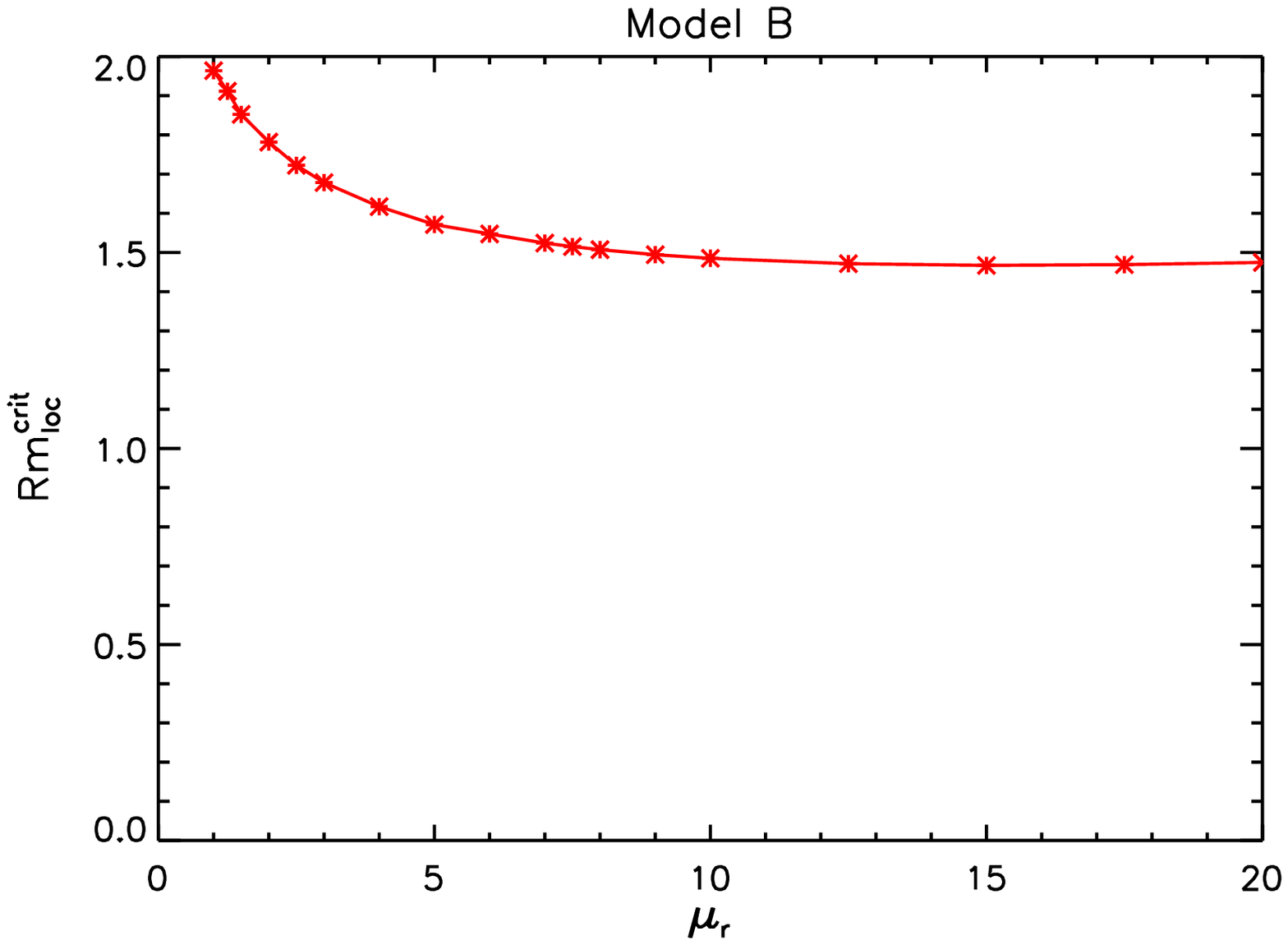}
\caption{Critical magnetic Reynolds number for the onset of dynamo
  action versus permeability. Left: flow model A with 16 helical cells, right:
  flow model B with 9 helical cells.
}\label{fig::rmcrit_vs_mur}
\end{figure}

Regarding the behavior of the growth-rates obtained with all relevant
mean-field coefficients, we find in general a good agreement between
FRM and MFM (solid and dashed curves in
figure~\ref{fig::mf_gr_vs_mur}g and~\ref{fig::mf_gr_vs_mur}h).
However, we see some increasing deviations at larger $\rml$ when
$\mur\ga 10$.  In that parameter regime the growth-rates obtained from
the MFM are systematically smaller than the growth-rates obtained from
the FRM.  The behavior of the growth-rates is not monotonic for model
A, whereas for model B we find an enhancement of induction action at
low $\mur$ while the growth-rates become independent of $\mur$ for
$\mur\ga 10$.  Considering the whole range of achievable $\mur$ in
model A we find a reduction of the critical magnetic Reynolds number
from ${\rm{Rm}}^{\rm{crit}}_{\rm{loc}}\approx 4.2$ (at $\mur=1$) to
${\rm{Rm}}^{\rm{crit}}_{\rm{loc}}\approx 3.2$ (at $\mur=20$).
However, inbetween, dynamo action is significantly suppressed by the
presence of ferromagnetic walls (left hand side in
figure~\ref{fig::rmcrit_vs_mur}) and $\rm{Rm}_{\rm{loc}}^{\rm{crit}}$
can even reach values up to $\sim 30$ around $\mur\approx 5.5$.
For model B we see a monotonic decrease from
${\rm{Rm}}_{\rm{loc}}^{\rm{crit}}\approx 2$ at $\mur=1$ to
${\rm{Rm}}_{\rm{loc}}^{\rm{crit}}\approx 1.5$ at $\mur=20$.  Regarding
the asymptotic behavior for large $\mur$ in
figure~\ref{fig::rmcrit_vs_mur} it seems unlikely that a further
increase of $\mur$ will significantly reduce the critical magnetic
Reynolds number of both flow models.

\section{Conclusions}

We have performed numerical simulations of the kinematic induction
equation for two different helical flow types including internal walls
or rods that may have magnetic properties.  In the limit of large
permeability, we found a moderate impact of $\mur$ on dynamo action in
terms of a reduction of ${\rm{Rm}}_{\rm{loc}}^{\rm{crit}}$ of roughly
25\% compared to the non-magnetic case.  This relative reduction of
the critical magnetic Reynolds number is nearly the same for both
models.  With view on the asymptotic behavior of
${\rm{Rm}}^{\rm{crit}}_{\rm{loc}}$ for large $\mur$ we do not expect
much smaller values for further increasing $\mur$.  
In model A, at the fluid-wall interface (where the field is maximum)
the magnetic field is predominantly parallel to the cell walls, so
that the permeability is not very important. The situation is less
clear for model B, for which one could guess that there is little
field in the rods because of flux expulsion from the helical flow, so
that the properties of the rods have little effect. 
Other possibilities for an explanation of the magnetic field
behavior rely on the particular topology of the permeability distribution,
which in our model B consists of disconnected columns. This might
hamper the formation of a large scale field, however,
the behavior is not unique in the whole parameter range so that more
detailed investigations are required to find a convincing explanation
for model B.
Regarding the
impact of the magnetic permeability, its influence on the critical
magnetic Reynolds number is less than what could have been guessed
from the results of VKS dynamo experiment.
This can be explained by the dominant dynamo mode which, in
the present study, can be characterized by the vertical
wavenumber. Here, the leading mode has the wavenumber $k_z=1$, so that
our results should be compared with the 
behavior of the simplest non-axisymmetric eigenmode in the VKS
configuration (the $m=1$ mode which is $\propto \cos\varphi$). Indeed,
\citeasnoun{2012NJPh...14e3005G} found a reduction of 29\% from
${\rm{Rm}}^{\rm{crit}}=76$ to ${\rm{Rm}}^{\rm{crit}}=54$ 
for $\mu_{\rm{r}}\rightarrow\infty$ which is rather close to the
reduction we obtained in our present calculations. However, both
models (VKS and helical flow models in the 
present study) are quite different so that this
accordance might be an accident. Regarding a dynamo mode with $k_z=0$ (which
corresponds to a uniform field in the vertical direction), we do not see such a
strong impact on its growth-rate as found for the axisymmetric dynamo
mode in the VKS model.

Despite the similar reduction of ${\rm{Rm}}_{\rm{loc}}^{\rm{crit}}$
for both models in the limit of large $\mur$, we find an entirely
different behavior for the corresponding mean-field coefficients.  For
model A, the presence of magnetic walls surrounding a single helical
flow cell results in a suppression of the $\alpha$-effect and a
transition to a negative $\beta$-effect (which remains smaller than
the ``normal'' diffusivity). In contrast, we see a slight enhancement of
$\alpha$ and a linear growth of $\beta$ for increasing $\mur$ in model
B, where the helical flow surrounds a magnetic rod.  The development
of $\alpha$ and $\beta$ is not sufficient to explain the constant
behavior of the growth-rates for large $\mur$ and $\rml$ where for
increasing $\mur$ we find a constant growth-rate, a constant
$\alpha$-effect but a linearly growing (positive) $\beta$-effect.
Hence, an additional dynamo supporting effect must be present in order
to compensate the increasing losses due to the $\beta$-effect. The
only possibility within our study is the $\delta_2$ effect, which
indeed becomes an important contribution in that parameter regime.

Comparing the growth-rates obtained from fully resolved models with
the corresponding mean-field models we found a good agreement between
both approaches for non-magnetic material ($\mur=1$) and for materials
with $\mur\la 20$.  The main reason for discrepancies at larger $\mur$
is the difficulty to estimate reliable values for the mean-field
coefficients and the occurrence of eigenmodes with larger vertical
wavenumber that are not included in our mean-field approach.

Our results can be adopted to large scale systems in which the flow
consists of tens of thousands of helical flow cells that cannot be
resolved in a direct numerical simulation.  The simplest way does not
need any information on mean-field coefficients and directly uses the
scaling found for the critical magnetic Reynolds number,
${\rm{Rm}}_{\rm{loc}}^{\rm{crit}} \propto \sqrt{D}$ in the limit of
small $D$ (which goes along with a large number of helical cells).
However, in order to model realistic systems it is necessary to
consider non-periodic (insulating) boundary conditions and the flow
outside of the core which essentially describes a large recirculation
cell\footnote[7]{The consideration of the recirculating flow has been
quite important for example for modelling of the Riga dynamo where
the reverse flow ensures that the dynamo instability sets in as an
absolute instability.}.  Such global models can hardly be modelled
in direct numerical simulations of the full set of magnetohydrodynamic
equations so it makes sense to model the magnetic induction due to the
helical small scale flow through the corresponding mean-field effects,
which in a global model only prevail in a limited region.  The main
contributions in such mean field models originate from the
$\alpha$-effect and the $\beta$-effect.  For non-magnetic internals we
have confirmed that the $\alpha$-effect can be expressed in terms of a
``universal'' function $\Phi$ that allows a conclusion on $\alpha$ for
larger systems when flow scale and flow amplitude are known.  In
combination with the $\beta$-effect which is roughly independent of
the flow scale and behaves $\propto \rml^2$ for small magnetic
Reynolds numbers this allows a modelling of systems that may consist
of tens of thousands of individual helical cells embedded into some
large flow structure.

Of course, for any specific sodium fast reactor a reliable estimate of
the dynamo effect would require further detailed knowledge, such as
the size of the core, the number of fuel rods contained therein, and
the total flow rate. In addition, the arrangement of the fuel rods
and, thus, the flow field is not as simple as it is assumed in our
idealized model.  For example, the fuel rods are packed much more
densely within an assembly with a hexagonal shape.  A more detailed
model in such a geometry would require a combination of our models A
and B in order to consider the small scale helical flow within an
assembly as well as the walls that separate individual assemblies.
Furthermore, it should be noted that the pitch angle which describes
the relation of vertical to horizontal flow may have an impact on
dynamo action. In the present study, this parameter is fixed to unity
assuming equipartition between horizontal and vertical flow whereas
realistic fast reactors are characterized by a dominant vertical flow.
Nevertheless, we believe that a consideration of these details will
only result in minor modifications to our findings and are therefore
of secondary importance.
 
\ack{We acknowledge support from the Helmholtz Alliance LIMTECH. We
  thank one anonymous referee for helpful comments and hints that
  helped to improve our mansucript.}

\section*{References}
\bibliographystyle{jphysicsB}

\end{document}